\documentclass{aa}

\usepackage{graphicx}
\usepackage{txfonts}
\usepackage{upgreek}
\usepackage{arydshln}
\usepackage{placeins}
\usepackage[dvipsnames]{xcolor}

\definecolor{simA}{HTML}{B2DF8A}
\definecolor{simB}{HTML}{ABCEE3}
\definecolor{simC}{HTML}{FFFF99}
\definecolor{simD}{HTML}{33A02C}
\definecolor{simE}{HTML}{1F78B4}
\definecolor{simF}{HTML}{000000}
\definecolor{simG}{HTML}{FDBF6F}
\definecolor{simH}{HTML}{1F78B4}
\definecolor{simI}{HTML}{33A02C}
\definecolor{simJ}{HTML}{6A3D9A}
\definecolor{simJt}{HTML}{CAB2D6}
\definecolor{simK}{HTML}{FDBF6F}
\definecolor{simL}{HTML}{B15928}
\definecolor{simM}{HTML}{A6CEE3}
\definecolor{simN}{HTML}{33A02C}
\definecolor{simNt}{HTML}{B2DF8A}
\definecolor{simO}{HTML}{1F78B4}
\definecolor{simOt}{HTML}{A6CEE3}
\definecolor{simP}{HTML}{A6CEE3}
\definecolor{simQ}{HTML}{000000}
\definecolor{simR}{HTML}{E31A1C}
\definecolor{simS}{HTML}{FB9A99}
\definecolor{simT}{HTML}{000000}
\definecolor{simU}{HTML}{B2DF8A}
\definecolor{simV}{HTML}{33A02C}
\definecolor{simW}{HTML}{1F78B4}
\definecolor{simX}{HTML}{A6CEE3}
\definecolor{simY}{HTML}{E31A1C}
\definecolor{simZ}{HTML}{FB9A99}

\begin{document} 

    \title{Accurately simulating core-collapse self-interacting dark matter halos}
    \titlerunning{Simulating core-collapse SIDM halos}

   \author{Moritz S.\ Fischer
          \inst{\ref{inst:usm},\ref{inst:origins}},
          Hai-Bo Yu\inst{\ref{inst:ucr}},
          Klaus Dolag\inst{\ref{inst:usm},\ref{inst:mpa}}
          }
    \authorrunning{M.\ S.\ Fischer, H.-B.\ Yu, K.\ Dolag}

    \institute{
        Universitäts-Sternwarte, Fakultät für Physik, Ludwig-Maximilians-Universität München, Scheinerstr.\ 1, D-81679 München, Germany\label{inst:usm}\\
        \email{mfischer@usm.lmu.de}
        \and
        Excellence Cluster ORIGINS, Boltzmannstrasse 2, D-85748 Garching, Germany\label{inst:origins}
        \and
        Department of Physics and Astronomy, University of California, Riverside, California 92521, USA\label{inst:ucr}
        \and
        Max-Planck-Institut f\"ur Astrophysik, Karl-Schwarzschild-Str. 1, D-85748 Garching, Germany\label{inst:mpa}
    }

   \date{Received 30 June, 2025 / Accepted 17 September, 2025}

  \abstract 
  {The properties of satellite halos provide a promising probe for dark matter (DM) physics. Observations have motivated current efforts to explain surprisingly compact DM halos. If DM is not collisionless, but has strong self-interactions, halos can undergo gravothermal collapse, leading to higher densities in the central region of the halo. However, it is challenging to model this collapse phase from first principles.}
  {To improve on this, we sought to better understand the numerical challenges and convergence properties of self-interacting dark matter (SIDM) $N$-body simulations in the collapse phase. Especially, our aim was to better understand the evolution of satellite halos.}
  {To do so, we ran SIDM $N$-body simulations of a low-mass halo in isolation and within an external gravitational potential. The simulation set-up was motivated by the perturber of the stellar stream GD-1.}
  {We find that the halo evolution is very sensitive to energy conservation errors, and a SIDM kernel size that is too large can artificially speed up the collapse.
  Moreover, we demonstrate that the King model can describe the density profile at small radii for the late stages that we have simulated.
  Furthermore, for our most highly resolved simulation ($N=5\times 10^7$) we have made the data public. It can serve as a benchmark.}
  {Overall, we find that the current numerical methods do not suffer from convergence problems in the late collapse phase and provide guidance on how to choose numerical parameters, for example that the energy conservation error is better kept well below 1\%.
  This allows simulations to be run of halos that become concentrated enough to explain observations of GD-1-like stellar streams or strong gravitational lensing systems.}
  

   \maketitle

\section{Introduction} \label{sec:introduction}

A plethora of observations including the clustering of galaxies and their rotation curves can be explained by assuming an invisible matter component, dark matter (DM). Within the cosmological standard model of Lambda cold dark matter ($\Lambda$CDM), this matter component mainly interacts gravitationally.
Numerous experiments have been conducted to discover DM via non-gravitational interactions with standard model particles. However, unambiguous evidence for such interactions is missing \citep[e.g.][]{Cirelli_2024}. Nevertheless, there are hints from astronomical observations that DM may have self-interactions, thus altering the evolution and structure of DM halos.

Self-interacting dark matter (SIDM) has historically been introduced to solve problems on galactic scales. In particular, it was meant to reduce the DM densities in the inner region of halos and to lower the abundance of satellites, i.e.\ suppress substructure \citep{Spergel_2000}. Even if those problems can be solved with baryonic physics and improved comparison to observations \citep[for reviews on small-scale problems of $\Lambda$CDM see][]{Bullock_2017, Sales_2022}, the underlying problem of understanding the nature of DM remains and poses one of the largest challenges in contemporary physics, thus making it imperative to search for signatures of DM physics.

There have been extensive studies on the impact of DM self-interactions on various astrophysical systems. This, for example, includes merging galaxy clusters \cite[e.g.][]{Robertson_2017b, Fischer_2023, Sabarish_2024, Valdarnini_2024} or dwarf galaxies \citep[e.g.][]{Vogelsberger_2014, Fry_2015, Ren_2019, Ebisu_2022, ManceraPina_2024, Kong_2025}.
Moreover, these systems have been used to constrain the strength of DM scattering as a function of the relative velocity of the particles \citep[][]{Kaplinghat_2016, Yang_2024}. For a comprehensive overview, we refer to the review articles by \cite{Tulin_2018, Adhikari_2022}.

The impact of DM self-interactions on the evolution of a DM halo can be described with the help of heat conduction \citep{Balberg_2002}. The scatterings among the DM particles lead to an energy transfer that follows the velocity dispersion gradient of the DM. This implies that a DM halo may experience heat conduction inwards to its centre when a positive velocity dispersion gradient is present. This is, for example, the case for a Navarro--Frenk--White (NFW) profile \citep{Navarro_1996}. As a consequence of the heat flow, the central density decreases.

When heat conduction continues, the velocity dispersion gradient becomes negative at all radii. This means that heat only flows outwards. While the central region of the halo is losing energy, it becomes more compact and the density increases. This phase of the halo evolution has been named the collapse phase. Here, the self-interactions lead to a runaway process resulting in gravothermal catastrophe \citep{Lynden-Bell_1968, Burkert_2000}.

Interestingly, some observations point towards more compact substructures, which could potentially be explained by the gravothermal collapse of SIDM halos \citep[e.g.][]{Turner_2021, Yang_2021, Yang_2023Da, Gad-Nasr_2024, Dutra_2024, Ragagnin_2024, Shah_2024, Zeng_2025a}. This includes gravitational lensing \citep[e.g.][]{Meneghetti_2020, Granata_2022}, in particular the observation of satellites acting as strong lens perturbers \citep[e.g.][]{Vegetti_2010, Enzi_2024, Despali_2025, Cao_2025, Li_2025, Minor_2025}.
In particular, \cite{Nadler_2023} found that the projected logarithmic density profile slope of their SIDM simulations for the collapse phase is consistent with the observations by \cite{Minor_2021}.
While gravitational lensing is very powerful in detecting substructures \citep[e.g.][]{Gilman_2021, Keeley_2024, Gannon_2025}, especially if they are dark, there are other possibilities to learn about substructures.

One of them is via their impact on stellar streams \citep[e.g.][]{Lu_2025}. For example, the stellar stream GD-1, which is one of the coldest and longest streams of the Milky Way, has a rich morphology \citep[e.g.][]{Grillmair_2006, Tavangar_2025}. It shows gaps \citep[e.g.][]{Carlberg_2013, deBoer_2018, deBoer_2020, Banik_2021, Malhan_2022} and a spur \citep[e.g.][]{Price-Whelan_2018, Bonaca_2018, Bonaca_2020}. These structures can be explained by compact perturbers passing by the stream.
Depending on the DM model, we expect compact dark substructures to be more or less abundant. \cite{Zhang_2025} showed that SIDM can form objects compact enough to explain the gaps and spur-like feature in GD-1, while CDM might be able to only explain the gaps.

The abundance and compactness of the substructures leave their distinct impact on stellar streams. Predictions for GD-1-like streams in the scenario of collisionless DM have been derived by \cite{Adams_2025}.
Additionally, self-interactions could directly impact the satellite from which the stellar stream originates. This is particularly relevant if the satellite is a dwarf galaxy in contrast to a star cluster. The self-interaction of the satellite's DM halo can potentially have at least a small impact on the morphology of the stellar stream \citep[e.g.][]{Zeng_2025b, Hainje_2025}.
For the future, we expect the wealth of observational data on stellar streams to increase strongly, not only for streams of the Milky Way \citep{Bonaca_2024}, but also for extragalactic stellar streams \citep[e.g.][]{Pearson_2022}. This opens up an opportunity to constrain DM models, especially SIDM and eventually find evidence for non-gravitational DM interactions.

Modelling DM substructures from first principles, in particular gravothermally collapsing SIDM satellites, is challenging \citep[e.g.][]{Yang_2022D, Zhong_2023, Mace_2024, Palubski_2024}.
Problems leading to energy non-conservation in $N$-body simulations of the late collapse phase of isolated halos have been explained in detail in \cite{Fischer_2024b}.
However, a small error in energy conservation does not necessarily imply that the simulation results are accurate. Checking the conservation of total energy can only serve as a diagnostic, but it does not guarantee reliable results.
Instead, it is necessary to run a simulation with high resolution and parameters tuned to achieve highly accurate results to obtain a benchmark against which other simulations can be tested.

Several aspects concerning the accuracy of simulations of the gravothermal collapse, such as their convergence behaviour, have not yet been fully investigated.
Therefore, for this paper, we studied the numerical properties of such SIDM simulations using the $N$-body code \textsc{OpenGadget3}.
In addition, we studied the qualitative differences arising from the velocity and angular dependence of the SIDM cross-section for a halo in isolation and as a satellite.
For the latter, our set-up was motivated by the GD-1 perturber.
As a result, we comment on how to choose numerical parameters for SIDM simulations. Moreover, we show that continuing an SIDM simulation with a too large time step may lead to incorrect predictions on the enclosed mass. Finally, we demonstrate that a King model provides a good description of the density profile at small radii in the collapse phase, and our results can be applied to study strong lensing perturbers.

The remainder of the paper is structured as follows.
In Sect.~\ref{sec:numerical_setup} we describe the numerical set-up of this study. The results of isolated halo simulations are presented in Sect.~\ref{sec:results1}, followed by Sect.~\ref{sec:results2} with a presentation of the findings when the halo is evolved in an external potential. In Sect.~\ref{sec:discussion} we discuss the limitations of our simulations as well as directions to continue this research. Finally, we summarise and conclude in Sect.~\ref{sec:conclusion}. Additional information is provided in the appendices.

\section{Numerical set-up} \label{sec:numerical_setup}

In this section we explain the employed numerical set-up. We begin with the simulation code and continue with the initial conditions (ICs) and finally summarise the simulation parameters.

For our simulations we used the cosmological $N$-body code \textsc{OpenGadget3}.
It is a successor of the \textsc{Gadget-2} code \citep{Springel_2005}.
The domain decomposition and the neighbour search we used have been described by \cite{Ragagnin_2016}. Additional information on the code is also given by \cite{Groth_2023} and the references therein.

\textsc{OpenGadget3} contains a module to simulate DM self-interactions, introduced by \cite{Fischer_2021a, Fischer_2021b, Fischer_2022, Fischer_2024a} and has been used for numerous studies of SIDM.
A unique feature of the SIDM module is that it is capable of simulating very anisotropic cross-sections \citep[see also][]{Fischer_2021a, Arido_2025}. Moreover, it allows one to perform the computations in parallel employing a message passing interface (MPI) parallelisation and/or an open multi-processing (OpenMP) parallelisation. At the same time, it explicitly conserves energy and linear momentum. A separate time step criterion was implemented for the self-interactions, which  ensures small-enough time steps. For computing the interactions between numerical particles, a spline kernel was employed \citep{Monaghan_1985}. The kernel size for each particle was chosen adaptively and set by the next $N_\mathrm{ngb} = 48$ neighbours if not stated otherwise.

We chose the ICs for our simulations motivated by the perturber of the stellar stream GD-1. It should be a halo that is a reasonable progenitor evolving to a state potentially explaining the observations. In this sense, a high concentration is desirable, as it makes rapid gravothermal evolution driven by SIDM more plausible.
Our ICs follow an NFW profile \citep{Navarro_1996}:
\begin{equation}
    \rho(r) = \frac{\rho_0}{\frac{r}{r_\mathrm{s}}\left(1+\frac{r}{r_\mathrm{s}}\right)^2} \,.
\end{equation}
We set the density parameter to $\rho_0 = 4.42 \times 10^{7} \, \mathrm{M_\odot} \, \mathrm{kpc}^{-3}$ and the scale radius to $r_\mathrm{s} = 1.28 \, \mathrm{kpc}$. The halo has a mass of about $\approx 2.8 \times 10^9 \, \mathrm{M_\odot}$ and its concentration parameter is $c \approx 22$, which is slightly higher than the typical concentration at this halo mass \citep{Dutton_2014}.
For our simulations, we generated the ICs using \textsc{SpherIC} \citep{Garrison_Kimmel_2013} and sampled the halo up to a radius $r_\mathrm{cut} = 10 \, r_\mathrm{s}$ or $r_\mathrm{cut} = 15 \, r_\mathrm{s}$. The ICs are all without baryons, i.e.\ DM only.

We set a spherical boundary condition at 200 kpc to avoid problems arising from particles travelling very far from the halo.
The boundary condition reflects particles that are travelling outwards, i.e.\ it does not directly affect energy conservation \citep[as done by][]{Fischer_2024b}.
The self-interactions locally thermalise the velocity distribution and consequently populate the high-velocity tail of the Maxwell-Boltzmann distribution. This leads to a few particles escaping the halo. When those particles travel to large radii, this can lead to significant round-off errors in the gravity calculations, given that we store the positions of the particles using double-precision floating-point numbers.
In our case, the precision of the particles may not be of much concern, given that we do not run our simulations for a very long time, and thus the particles may not travel to very large distances. However, to ensure that no problem arises from a reduced precision in the position of the particles, we introduced the boundary condition.

Moreover, we simulated the halo with three different mass resolutions and also varied the gravitational softening length. The most highly resolved simulations contain $N=5 \times 10^7$ particles. Furthermore, we employed different values for the gravitational time step parameter $\eta$ \citep[eq.~34 by][]{Springel_2005} or the SIDM time step parameter $\tau$ \citep[eq.~5 by][]{Fischer_2024a} and employed for some simulations a minimum allowed time step $\Delta t_\mathrm{min}$. In addition, we set the opening angle for the gravitational force computations by setting the value of the parameter $\alpha=5\times10^{-4}$ \citep[eq.~18 by][]{Springel_2005}. An overview with the details of our simulations is provided in Table~\ref{tab:sim_para}.

For most of our simulations we assumed an isotropic velocity-independent cross-section. Simulations with a different angular or velocity dependence are described in Sect.~\ref{sec:angular_and_velocity_dep_iso}.
In addition to the isolated halo simulations, we placed the halo in an external potential. This allowed us to study its evolution when being subject to tidal forces. The details are described in Sect.~\ref{sec:satellite_setup}.
An overview of the 29 simulations that we show in this paper is given in Appendix~\ref{sec:sim_overview}. The choices of the various numerical parameters, such as the softening length or the values for the time step criteria, are highlighted.

\section{Isolated halo evolution} \label{sec:results1}

In this section we present our simulation results and discuss the relevance of various numerical parameters for achieving accurate simulations. This includes various aspects such as the role of variable time steps, the radius up to which the ICs are sampled, the role of mass resolution, the gravitational softening length and the SIDM kernel size, as well as setting a minimal time step. We also present results for different angular and velocity dependences and comment on their qualitative differences. Furthermore, we explore the evolution of the velocity anisotropy. In addition, we discuss how much the evolution of an SIDM halo simulation depends on the choice for the gravitational time step criterion in Appendix~\ref{sec:grav_time_step}. We also demonstrate in Appendix~\ref{sec:king_fit} that the inner region during the collapse phase can be well described by a King model \citep{King_1962}.

For all simulations, we used the peak find algorithm introduced by \cite{Fischer_2021b} to find the centre of the halo. It corresponds to the minimum of the gravitational potential. We computed physical quantities such as the enclosed mass with respect to this centre.

\subsection{Relation between softening length and variable time steps} \label{sec:soft_tstep}

\begin{figure}
    \centering
    \includegraphics[width=\columnwidth]{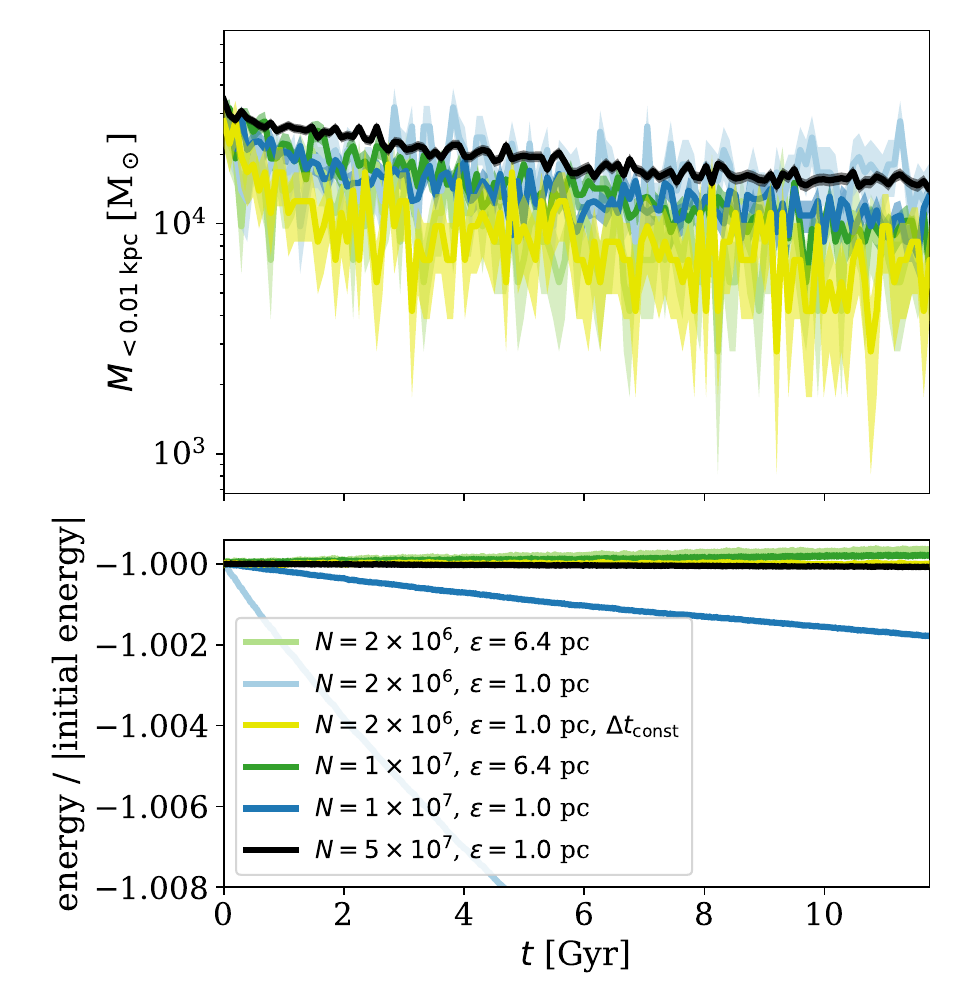}
    \caption{Interplay of softening length and adaptive time steps for energy conservation. We show the time evolution of the mass enclosed within 10 pc (upper panel) and the energy conservation (lower panel) for simulations of collisionless DM. The shaded regions indicate the uncertainties estimated based on shot noise. The simulations employ an adaptive time step, except for the results given by the yellow curve. More details can be found in Table~\ref{tab:sim_para}; the corresponding names are A, B, C, D, E, and F.}
    \label{fig:soft_tstep}
\end{figure}

Here we investigate the size of the gravitational softening length and its relation to the use of an adaptive time-stepping scheme. For this purpose, we consider only CDM simulations.

In Fig.~\ref{fig:soft_tstep} we can see that a too small gravitational softening length can lead to a loss of energy over the course of the simulation (compare light green and light blue). However, when fixing the adaptive time steps to a constant value for all particles over the whole simulation (yellow curve), the error in energy conservation vanishes.
This can be understood by a too small softening length, which causes more frequent changes in time steps, thus implying a larger error in total energy. The energy error related to the time steps is a consequence of choosing the time steps in a non-time-symmetric way \citep{Dehnen_2017}.

We note that we have chosen a fairly small time step for the simulation with the fixed time step.
This ensures that all particles are evolved on a sufficiently small time step, but at the same time reduces integration errors that are not directly related to the change in time steps as well.
If we instead choose a much larger constant time step for all particles, this leads to an increase in total energy by amplifying the errors that arise from the asymmetric tree evaluation that we do in \textsc{OpenGadget3}, implying that it decreases when shrinking the opening angle for the tree nodes \cite[see also the explanation given by][at the end of their Sect.~3.2.1]{Fischer_2024b}.
In addition, we also checked that the energy conservation error for the simulation with the small softening length and the variable time step (light blue) does not arise from the error related to the asymmetric tree evaluation, i.e.\ we explicitly checked that decreasing the opening angle does not affect the error in energy conservation.
In conclusion, the statement from the previous paragraph about the energy conservation error arising from the asymmetric change of time steps is valid.

\subsection{Sampling radius of ICs}

\begin{figure}
    \centering
    \includegraphics[width=\columnwidth]{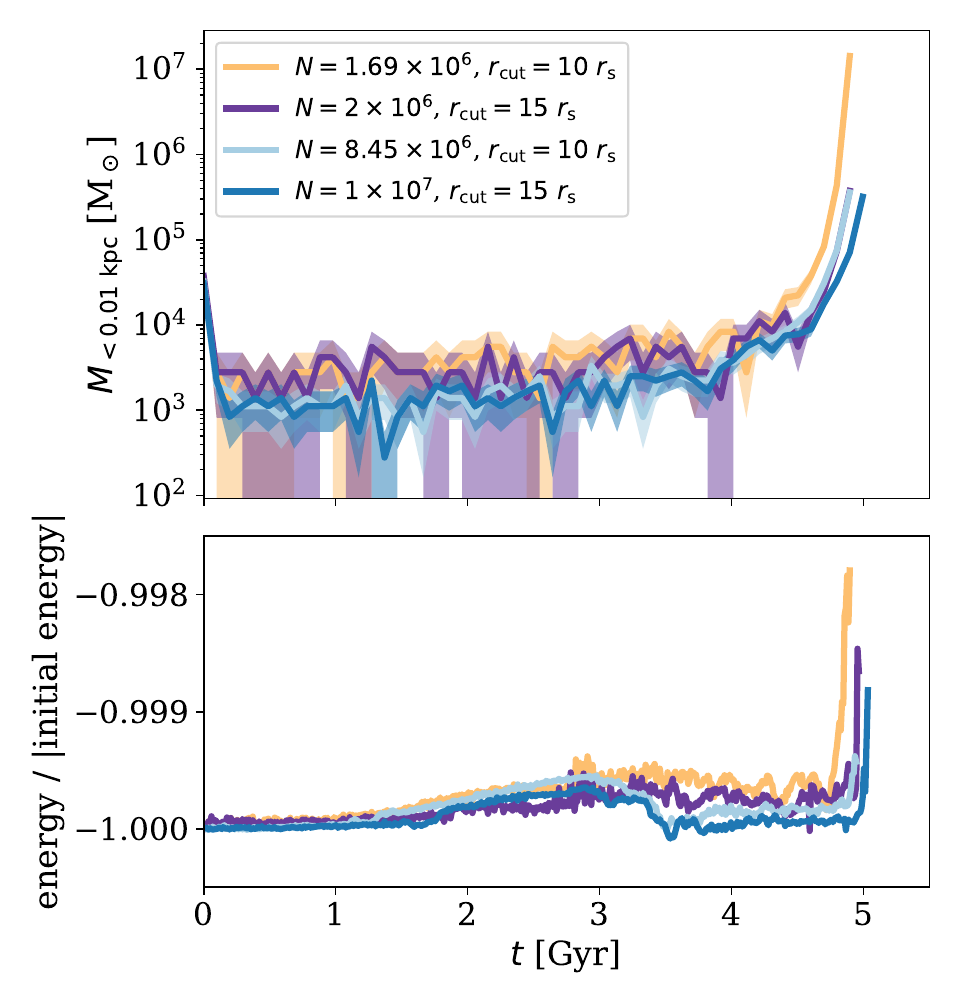}
    \caption{Variation of the maximum sampling radius for the ICs. Analogously to Fig.~\ref{fig:soft_tstep}, the mass enclosed within 10 pc (upper panel) and the energy conservation (lower panel) are shown as a function of time.
    The results for ICs sampled up to $10 \, r_\mathrm{s}$ (orange and light blue) and $15 \, r_\mathrm{s}$ (purple and dark blue) are shown.
    We note that the lower resolution simulations (orange and purple) employ a softening length of $\epsilon=6.4$ pc, whereas the higher resolution simulations (light and dark blue) employ a softening length of $\epsilon=1.0$ pc. All parameters for the shown simulations (G, J, M, and O) are given in Table~\ref{tab:sim_para}.}
    \label{fig:cut_rad}
\end{figure}

As a next step, we investigated the role of the ICs. In particular, we made an approximation that is commonly made when sampling a halo model such as the NFW profile.

Sampling the ICs for a density profile that implies an infinite mass, such as the NFW profile, requires truncating the halo at a radius $r_\mathrm{cut}$. The truncation of the halo can influence its gravothermal evolution and, for example, can speed up the collapse for satellite galaxies, where the outer regions of the halo are removed due to tidal force \citep[e.g.][]{Nishikawa_2020, Sameie_2020b}. In contrast to this physical truncation, we investigated the role of the artificially set truncation radius for SIDM simulations.

In Fig.~\ref{fig:cut_rad} we test the influence of $r_\mathrm{cut}$ on the evolution of the halo. A comparison of simulations with ICs sampled up to $10 \, r_\mathrm{s}$ and $15 \, r_\mathrm{s}$ reveals a significant deviation in the evolution.
Although the simulations agree with each other during the core formation and early collapse phase within the estimated Poisson noise based on the number of simulation particles within 10 pc, they start to deviate at later times. 
Here the enclosed mass grows super-exponentially, and the relative uncertainty on the inferred mass goes down.
At the same time, the difference between the simulations in terms of the enclosed mass becomes relatively large, although the deviation in the collapse time is small. Because of the super-exponential nature of the enclosed mass growth in the deep core-collapse phase, a small deviation in the collapse time results in a large difference in the mass. Depending on how accurately the evolution should be predicted, $r_\mathrm{cut} = 10 \, r_\mathrm{s}$ could be too small, and a higher value would be preferable.

Finally, we note that we chose $r_\mathrm{cut}$ in units of $r_\mathrm{s}$ compared to the virial radius. This is because the only relevant length scale for the NFW profile is $r_\mathrm{s}$, whereas the virial radius is only defined by making additional assumptions about cosmology, namely the critical density.

\subsection{The role of resolution and softening length}
\label{sec:resolution_and_softening}

\begin{figure}
    \centering
    \includegraphics[width=\columnwidth]{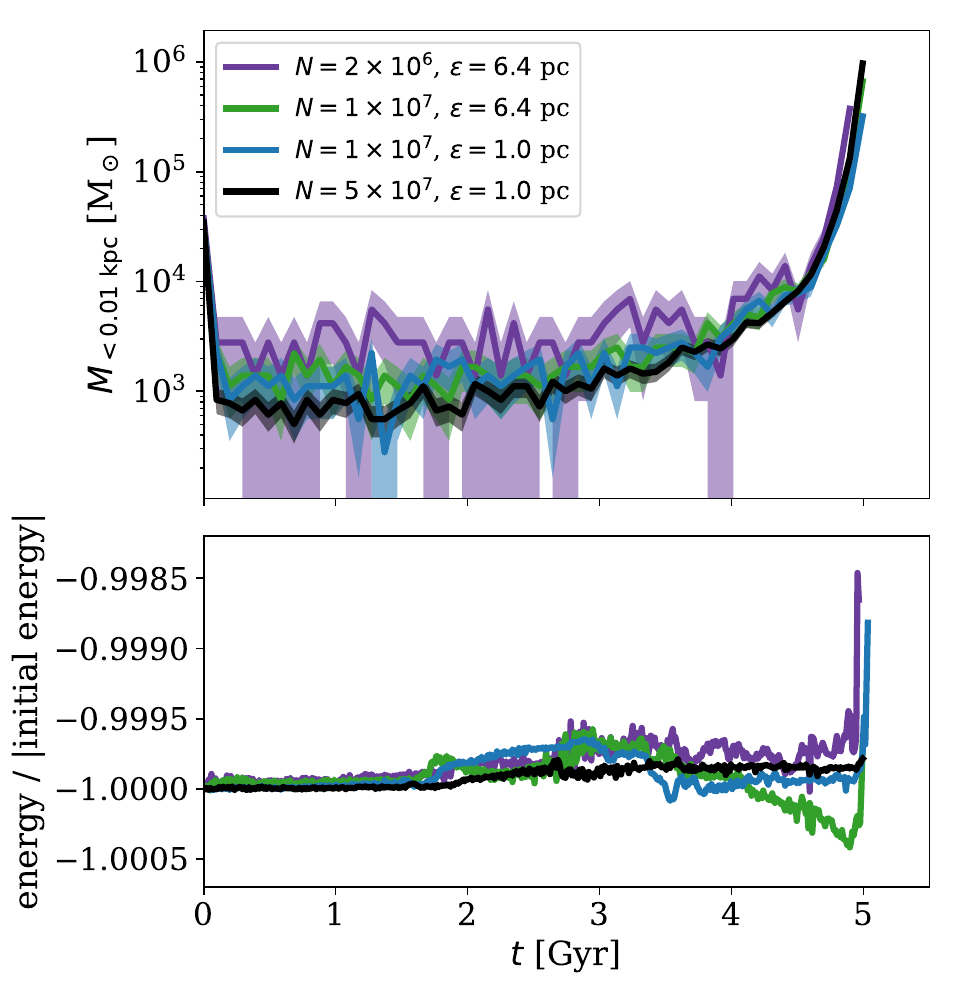}
    \caption{Variation of resolution and gravitational softening length. Similar to Fig.~\ref{fig:soft_tstep}, we show the enclosed mass within 10 pc and the energy conservation as a function of time. The simulations J, N, O, and Q as given in Table~\ref{tab:sim_para} are shown.}
    \label{fig:resolution+softening}
\end{figure}
We now look at the role of the numerical resolution, i.e.\ the number of numerical particles that are used to resolve the DM halo. At the same time, we consider two different choices for the gravitational softening length.

The results for four of our SIDM simulations are given in Fig.~\ref{fig:resolution+softening}.
Interestingly, from a comparison of the purple and black curves, we can see that the collapse time is increasing slightly with higher resolution. An effect that could contribute to this is gravitational two-body relaxation being stronger in simulations with lower resolution. An increased effective heat conduction due to a large SIDM kernel size compared to the mean free path can also contribute to this, as we discuss in Sect.~\ref{sec:kernel_size}.
We note that the difference between these two runs may not be caused by inaccurate energy conservation. In fact, the simulation indicated by the purple curve is slightly gaining energy, which would slow down the evolution (see also Sect.~\ref{sec:energy_conservation}). However, this is in contrast to what we find when we compare it with the black curve.

Moreover, we do not find clear evidence that the gravitational softening length has a significant impact on the evolution for the values that we have tested. This suggests that the results are not very sensitive to gravitational softening as long as the softening length is chosen within a reasonable range. The very slight difference between the green and blue curves for the enclosed mass might be related to differences in energy conservation.

A common choice is to set the gravitational softening length according to \cite{Power_2003} or \cite{van_den_Bosch_2018}. However, the optimal gravitational softening length for an SIDM halo in the collapse phase might be smaller than that for the pre-collapsing halo (e.g.\ an NFW halo). This is especially the case if one is interested in the inner regions of the collapsing object, i.e.\ fairly small length scales. As a consequence, as a compromise one may eventually want to choose a softening length that is somewhat smaller, closer to undersoftening than recommended according to the criteria based on CDM simulations.

\subsection{The role of energy conservation}
\label{sec:energy_conservation}

\begin{figure}
    \centering
    \includegraphics[width=\columnwidth]{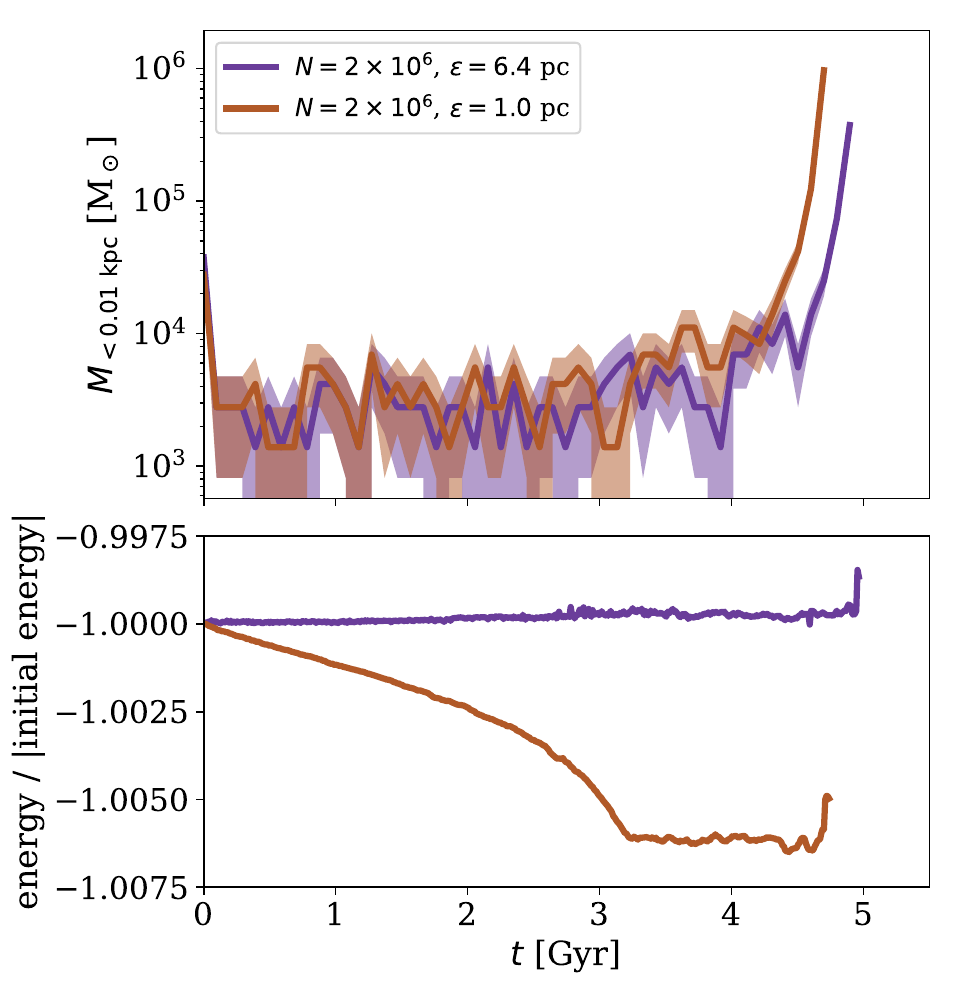}
    \caption{Impact of energy conservation on the evolution time of the halo. Following Fig.~\ref{fig:soft_tstep}, we show the enclosed mass within 10 pc (upper panel) and the energy conservation (lower panel) as a function of time. The shown simulations differ in softening length, causing differences in the accuracy of energy conservation (see explanations in Sect.~\ref{sec:soft_tstep}). All parameters for the displayed simulations J and L are given in Table~\ref{tab:sim_para}.}
    \label{fig:energy_cons}
\end{figure}

Several numerical difficulties can affect the conservation of the total energy over time. Next, we comment on its role in simulations of SIDM halos.

The gravothermal evolution leads to a collapse because the SIDM halo is losing energy in its central region, making DM sink further inside. Similarly, the gravothermal evolution can be impacted by numerical errors in the conservation of total energy. This may imply that a loss of total energy speeds up the collapse while an increase slows it down.

In Fig.~\ref{fig:energy_cons} we show two simulations that differ in energy conservation. The one with the smaller softening length (red) loses energy in line with the explanation given in Sect.~\ref{sec:soft_tstep}. It loses approximately 0.6\% of its total energy, resulting in a speed-up of the evolution of about 4.2\%, when compared to the simulation with the larger softening length (purple). We note that the larger softening length is not directly responsible for a significant slowdown of the collapse for the corresponding simulation (purple), as we found in Sect.~\ref{sec:resolution_and_softening}. Overall, this illustrates that an accurate prediction of the collapse time requires a fairly accurate conservation of total energy.

\subsection{SIDM kernel size}
\label{sec:kernel_size}

\begin{figure}
    \centering
    \includegraphics[width=\columnwidth]{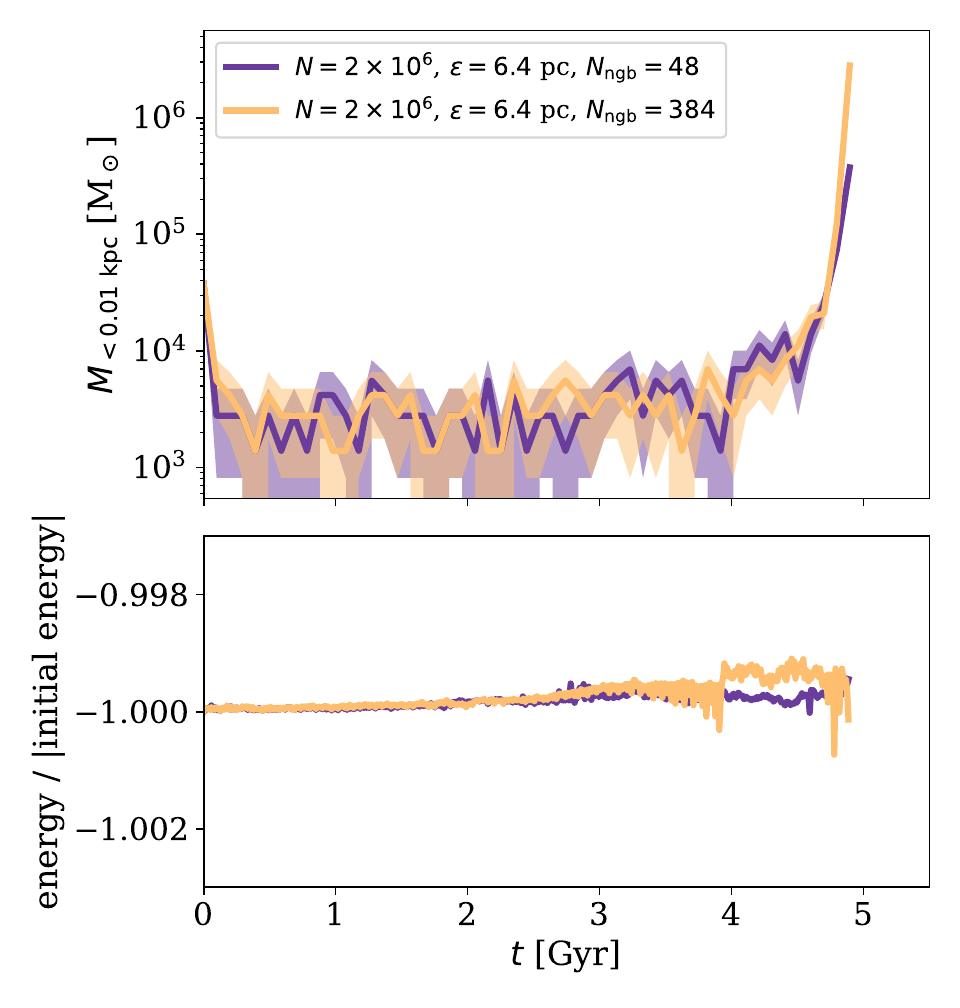}
    \caption{Evolution of an isolated halo as a function of time for two different choices of $N_\mathrm{ngb}$. The upper panel gives the mass enclosed within $10 \, \mathrm{pc}$ and the lower panel displays the energy conservation as a function of time. The choice of $N_\mathrm{ngb} = 384$ corresponds to twice the kernel size compared to the SIDM computations for $N_\mathrm{ngb} = 48$. Table~\ref{tab:sim_para} gives the parameters employed for the displayed simulations J and K.}
    \label{fig:Nngb}
\end{figure}

\begin{figure}
    \centering
    \includegraphics[width=\columnwidth]{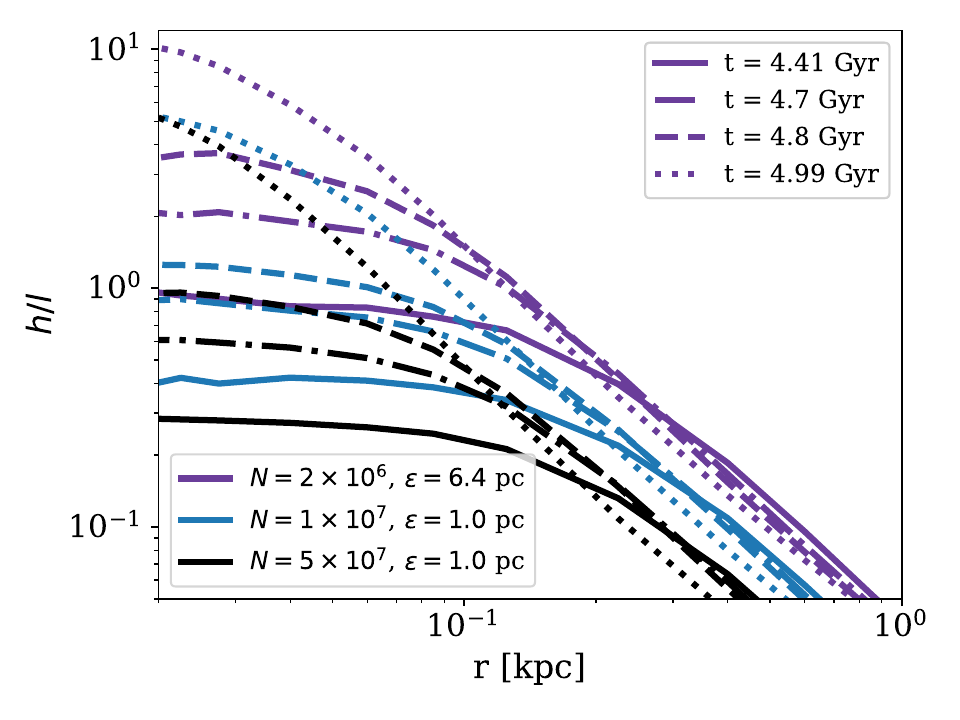}
    \caption{SIDM kernel size divided by the mean free path as a function of radius for several times. The results, computed according to Eq.~\eqref{eq:hDl}, are shown for several simulations with different resolutions. All simulations were run with $N_\mathrm{ngb} = 48$. More details for the shown simulations J, O, and Q are given in Table~\ref{tab:sim_para}.}
    \label{fig:hDl}
\end{figure}

It may have been \cite{Koda_2011} who first pointed out that SIDM $N$-body simulations should resolve the mean free path set by the local density and cross-section. \cite{Fischer_2024c} found for a set-up rather different from a halo simulation that the ratio of the SIDM kernel size, $h$, to the mean free path, $l$, actually matters. Here we investigate the relevance of this for simulating the collapse phase of SIDM halos. 

For this purpose, we reran one of our simulations, increasing the neighbour number from $N_\mathrm{ngb}=48$ to $N_\mathrm{ngb}=384$. This implies an increase in kernel size by a factor of two. In Fig.~\ref{fig:Nngb}, we show the results for the mass enclosed within 10 pc and the energy conservation. For almost all of the evolution, there is basically no difference between the two simulations. Only at very late stages, where $h/l$ has increased a great deal at small radii due to the collapse, do we find that the simulation with the larger value for $N_\mathrm{ngb}$ reaches high densities earlier. This could be explained by a larger SIDM kernel size leading to an artificially strong effective heat conduction.

We computed- the ratio of kernel size, $h$, to mean free path, $l$, following \cite{Fischer_2024b}:
\begin{equation} \label{eq:hDl}
\frac{h}{l} = \frac{\sigma_\mathrm{eff}}{m} \, \rho^{2/3} \, \sqrt[3]{\frac{3 \, m_\mathrm{n} \, N_\mathrm{ngb}}{\uppi \sqrt{2}}} \,.
\end{equation}
Here, $\sigma_\mathrm{eff}$ is defined as the effective cross-section given by Eq.~\ref{eq:effective_cross_section}. In the case of a velocity-independent isotropic cross-section, $\sigma_\mathrm{eff} = \sigma$.
The result is given for various simulations and times as a function of radius in Fig.~\ref{fig:hDl}. We find that a kernel size that is even a few multiples larger than the mean free path appears to be unproblematic (compare also Fig.~\ref{fig:resolution+softening}). In contrast, ratios exceeding an order of magnitude appear to be problematic.
However, in general, it depends on the simulation set-up how problematic a specific value for $h/l$ is; in particular, it should depend on the velocity dispersion gradient.
In this sense, stages of the gravothermal collapse later than those that we have simulated could be even more problematic.
Furthermore, we want to add that there might be a mild dependence on how the geometric factor $\Lambda$ for the interaction probability and the drag force is computed \citep[see eqs.~9, 13 and B4 by][]{Fischer_2021a}; for example, the chosen kernel function may matter. 

Finally, we note that the error arising from a too large value for $h/l$ does not show up in the energy conservation of the simulation. Hence, this is a good example of where energy conservation alone is not enough to assess the simulation's accuracy.

\subsection{Using a minimal time step}
\begin{figure}
    \centering
    \includegraphics[width=\columnwidth]{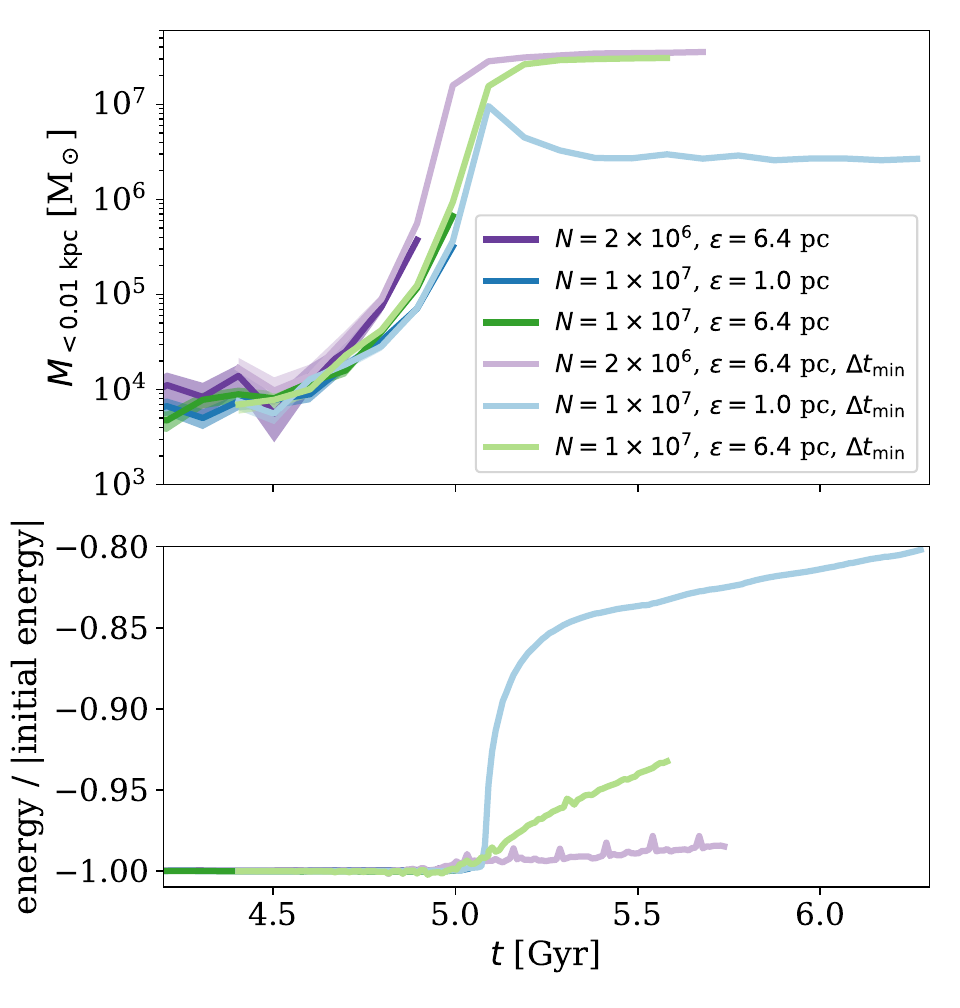}
    \caption{Employing a minimum time step. As in Fig.~\ref{fig:soft_tstep}, we show the enclosed mass within 10 pc and the energy conservation for various simulations. This time, we investigate how much the simulations are affected when employing a minimum time step. The brighter lines indicate simulations without a minimum time step, i.e.\ the time step can decrease further. In contrast the darker lines feature a minimum time step of $\Delta t_\mathrm{min} = 2.8 \times 10^{-5} \, \mathrm{Gyr}$. The displayed simulations J, N, O, Jt, Nt, and Ot are described in Table~\ref{tab:sim_para}.}
    \label{fig:min_tstep}
\end{figure}

\begin{figure}
    \centering
    \includegraphics[width=\columnwidth]{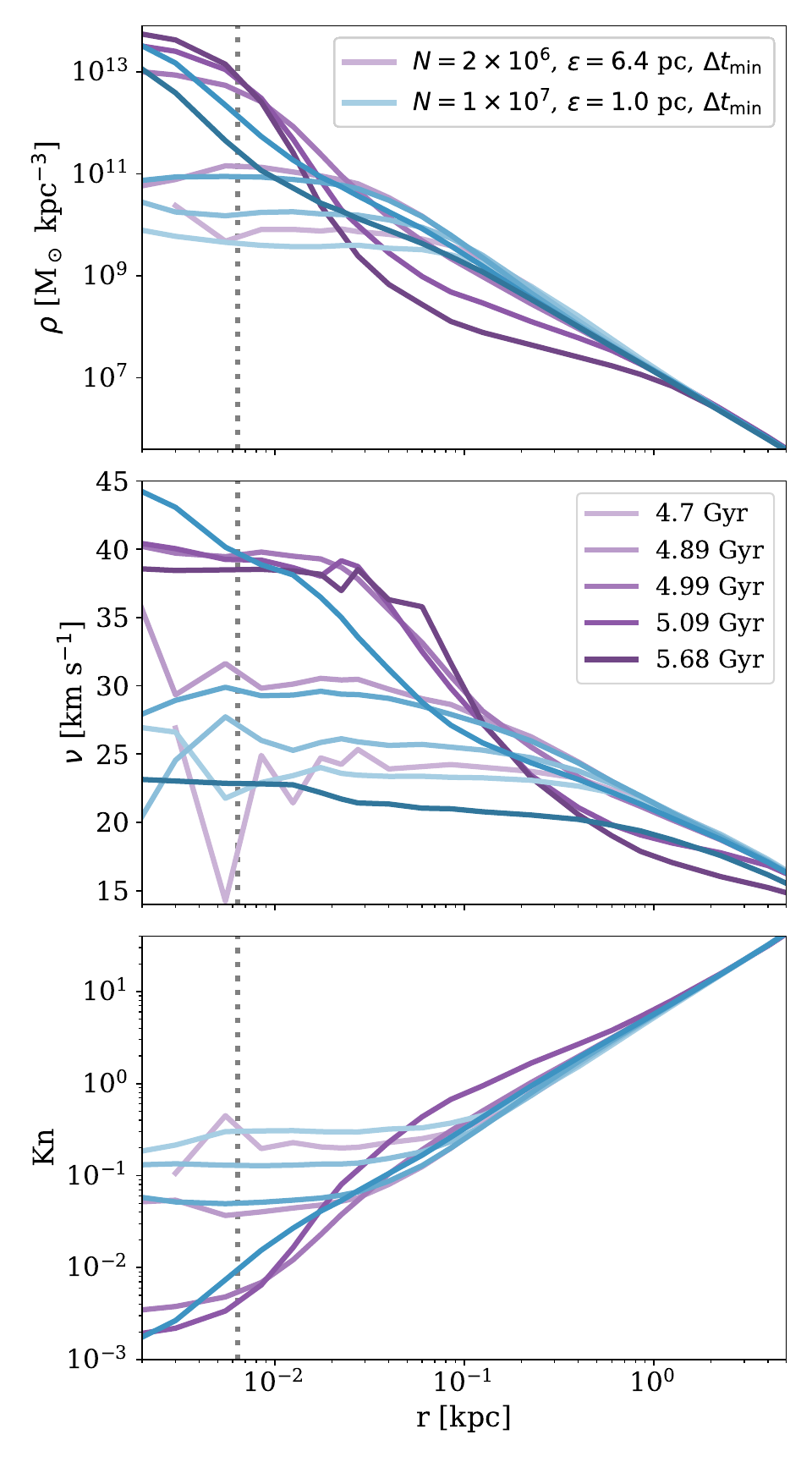}
    \caption{Density, velocity dispersion, and Knudsen number for simulations employing a minimum time step. The purple lines give the results for the simulation with $N=2\times 10^6$ and $\epsilon=6.4 \, \mathrm{pc}$ (grey) and $\Delta t_\mathrm{min} = 2.8 \times 10^{-5} \, \mathrm{Gyr}$ and the blue lines are for the simulation with $N=1\times 10^7$ and $\epsilon=1.0 \, \mathrm{pc}$ and $\Delta t_\mathrm{min} = 2.8 \times 10^{-5} \, \mathrm{Gyr}$. These are the same simulations as in Fig.~\ref{fig:min_tstep}. The top panel gives the density as a function of radius for different times. We note that these times are during the collapse phase, and for most of them the energy conservation error is sizeable. The middle panel gives the velocity dispersion as a function of radius, and the bottom panel displays the Knudsen number (Eq.~\eqref{eq:knudsen_number}). All parameters for the shown simulations Jt and Ot are given in Table~\ref{tab:sim_para}.}
    \label{fig:min_tstep_profiles}
\end{figure}

Next, we discuss how imposing a minimum time step impacts the simulation results in the collapse phase. 
The time step criterion for SIDM, but also gravity, requires smaller and smaller time steps while the halo is collapsing. In particular, the SIDM time step criterion becomes prohibitively small. This poses a challenge for cosmological simulations as one aims to run a simulation until a specific redshift (e.g.\ $z=0$). A strategy to complete such a simulation, although some halos are collapsing, is to set a minimum allowed time step despite the time step criteria, i.e.\ evolving some particles on too large time steps.

Following this idea, we set a minimum time step for the late evolution phase. In practice, we continued some of our simulations with this minimum time step. In our case, we chose $\Delta t_\mathrm{min} = 2.8 \times 10^{-5} \, \mathrm{Gyr}$.
The results of the simulations for which we tested this are displayed in Fig.~\ref{fig:min_tstep}. Here we show the mass enclosed within 10 pc and the energy conservation error.
For the late collapse phase, we find that the simulations with the minimum time step agree well with those that can contain particles on even smaller time steps. Having a minimum time step in place, we were able to continue the simulations to even later stages with higher central densities. However, at some point, the central density stopped growing, and the total energy was no longer conserved, but increased drastically. 

Interestingly, the increase in total energy seems to be sensitive to the gravitational softening length and number of particles. For the three runs with the fixed minimum time step, the one with the highest number of particles and the smallest softening length actually has the most severe energy violation. The cause of this surprising behaviour is not entirely clear. Although we impose a minimum time step, the size of the time step at which a particle is evolved can still change. As a consequence, we expect a contribution from the non-time-symmetric change of time steps to the energy conservation error. We note that this error can depend on the choice of the gravitational softening length (see Sect.~\ref{sec:soft_tstep}). Another source of the energy conservation error comes from the asymmetry of the one-sided oct-tree used in the gravitational force computations \citep{Fischer_2024b}. In addition, how quickly the numerical errors grow depends on the size of the employed time step. In particular, the ratio of the actually used time step to the one implied by the time step criteria can matter, and are independent of the adopted minimum time step. Thus, although the minimum time step is fixed to a constant, the time step required by the time step criteria depends effectively on the number of particles and the softening length. The interplay of these factors complicates the evolution of the energy conservation error. 

We would like to emphasise that although the runs with relatively low resolution have better energy conservation, it does not mean they are accurate. For example, the simulations shown reach densities for which the SIDM kernel sizes might be relatively large compared to the mean free path, leading to an underestimation of the collapse time, as discussed in Sect.~\ref{sec:kernel_size}. To efficiently and accurately simulate such late stages, more work is needed, especially in developing measures to evaluate the accuracy of the simulations, as well as alternative models to describe the inner region of the halo, as we discuss in Sect.~\ref{sec:discussion}.

Studying the mass enclosed only within 10 pc may give an incomplete picture of the impact of imposing a minimum time step. Instead, we consider the density, the velocity dispersion and the Knudsen number,
\begin{equation} \label{eq:knudsen_number}
    Kn = \sqrt{\frac{4 \uppi \, \mathrm{G}}{\rho \, \nu^2}} \left(\frac{\sigma_\mathrm{eff}}{m}\right)^{-1} \,.
\end{equation}
We use the effective cross-section, $\sigma_\mathrm{eff}$, given by Eq.~\ref{eq:effective_cross_section}. Moreover, for a velocity-independent isotropic cross-section, $\sigma_\mathrm{eff} = \sigma$.
We computed those quantities as a function of radius and show them for two of our continued simulations in Fig.~\ref{fig:min_tstep_profiles}.
For the first time that we show ($4.7 \, \mathrm{Gyr}$), the two simulations give fairly similar results. However, at later times they deviate from each other, sometimes leading to huge differences (e.g.\ for the velocity dispersion).
This demonstrates that measuring quantities from simulations of collapsing halos where a minimum time step had been imposed can be far off.

Nevertheless, the simulation Jt shown in purple in Fig.~\ref{fig:min_tstep_profiles} might be more accurate as the energy conservation is much better. The results for the very late collapse phase, where we find an increase in the central density compared to the maximum core formation by a factor of $6 \times 10^5$, also appear to be qualitatively in agreement with the findings in Fig.~3 by \cite{Balberg_2002} for the gravothermal fluid model.

In Sect.~\ref{sec:projected_enclosed_mass}, we comment further on the errors arising from imposing a minimum time step. In detail, we look at the enclosed mass in projection, which is relevant for gravitational lensing.

\subsection{Evolution of projected enclosed mass}
\label{sec:projected_enclosed_mass}

\begin{figure}
    \centering
    \includegraphics[width=\columnwidth]{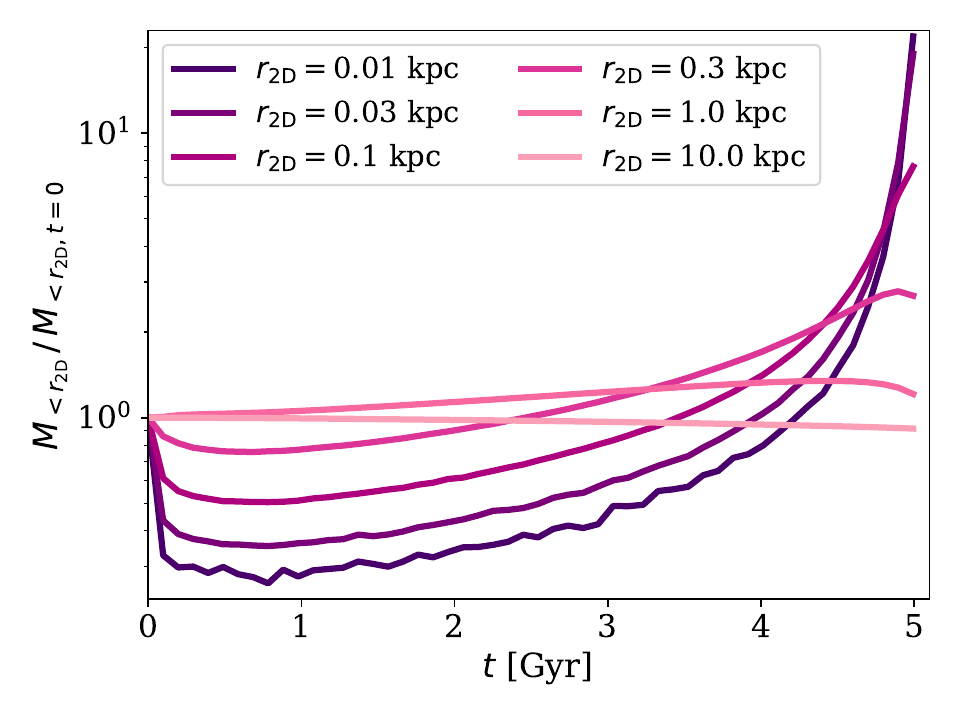}
    \caption{Time evolution of the enclosed mass for our most highly resolved simulation (simulation Q in Table~\ref{tab:sim_para}). The relative change of the enclosed mass within various radii is shown as a function of time. The enclosed mass is computed in projection, i.e.\ it is the mass within a cylinder of radius $r_\mathrm{2D}$.}
    \label{fig:enclosed_mass}
\end{figure}

We wanted to get closer to the observations, in particular gravitational lensing, and to measure the enclosed mass in projection.
In Fig.~\ref{fig:enclosed_mass}, the enclosed mass within several radii is shown as a function of time for our most highly resolved simulation ($N=5 \times 10^7$).
For small radii, we find that the enclosed mass first decreases during the core expansion phase and later increases when the gravothermal collapse dominates, as we also found in our three-dimensional analysis. This is in contrast to the bound mass of the system. The self-interactions create a few particles that have velocities exceeding the escape velocity, leading to a decreasing mass of the halo. Even more particles may gain enough kinetic energy to travel to fairly large radii, but still remain bound. In line with this, we find that for the largest radius shown in Fig.~\ref{fig:enclosed_mass}, the mass decreases slightly.
Related to this, the velocity anisotropy at large radii increases, i.e.\ radial motion becomes more important, as we show in Appendix~\ref{sec:high_res}.

If one defines a radius $r_\mathrm{eq}$ as the one at which the mass inflow and the mass outflow have the same size, i.e.\ $\dot{M}(<r_\mathrm{eq}) = 0$. Then $r_\mathrm{eq}$ decreases with time when the halo is undergoing the gravothermal collapse. This implies that we do not expect the mass within a fixed radius (e.g.\ 1 kpc) to keep continuously increasing. Instead, it starts to decrease as we can see for a few of the radii that we have chosen for Fig.~\ref{fig:enclosed_mass}.
The smaller the radius, the later the point in time where the enclosed mass starts to decrease.\footnote{For a more complicated model, such as the one studied by \cite{Patil_2025}, the decrease in mass in the late stages can be avoided by approaching a stable solution with an increased enclosed mass compared to collisionless DM instead of undergoing a gravothermal catastrophe.}

\begin{figure}
    \centering
    \includegraphics[width=\columnwidth]{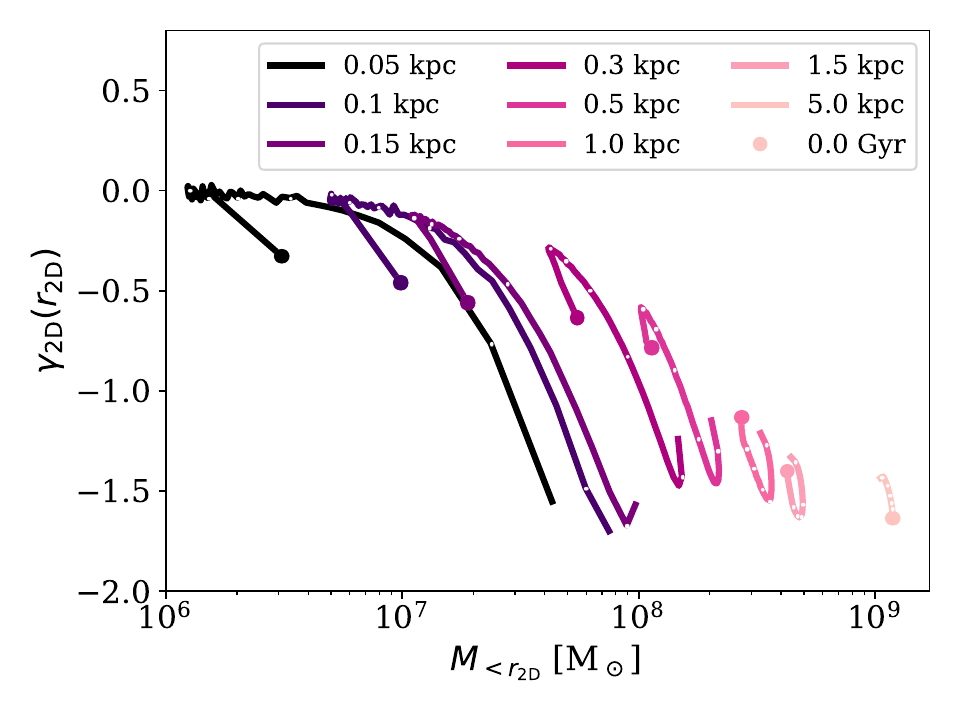}
    \caption{Projected logarithmic density profile slope as a function of projected enclosed mass. For our most highly resolved simulation evolved in isolation (simulation Q from Table~\ref{tab:sim_para}), we show how the projected logarithmic density profile slope, $\gamma_\mathrm{2D}$, and the projected enclose mass, $M_{<r_\mathrm{2D}}$, evolve with time. We display $\gamma_\mathrm{2D}$ and $M_{<r_\mathrm{2D}}$ for several radii $r_\mathrm{2D}$ (see inset for legend). The larger circles mark the value at the beginning of the simulation, i.e.\ for an NFW profile. Moreover, the small white dots are placed equidistant in time, every 0.98 Gyr.}
    \label{fig:logslope_vs_mass}
\end{figure}
The projected logarithmic density profile slope, $\gamma_\mathrm{2D}$, shows a similar behaviour.
In Fig.~\ref{fig:logslope_vs_mass}, we give $\gamma_\mathrm{2D}$ as a function of the enclosed mass $M_{<r_\mathrm{2D}}$. The two quantities can be inferred from observations via gravitational lensing analysis. Here we use our most highly resolved simulation, which models the halo in isolation. The figure shows how the system evolves in the $\gamma_\mathrm{2D}$--$M_{<r_\mathrm{2D}}$ plane considering several radii, $r_\mathrm{2D}$. It can be seen that not only does the enclosed mass make a turn during the collapse phase, as discussed above, but the density slope also starts to become flatter at a specific point in time. Similarly to the enclosed mass, this turn occurs earlier at larger radii. Interestingly, $\gamma_\mathrm{2D}$ turns even earlier to becoming flatter than the projected enclosed mass starts to decrease.

\begin{figure}
    \centering
    \includegraphics[width=\columnwidth]{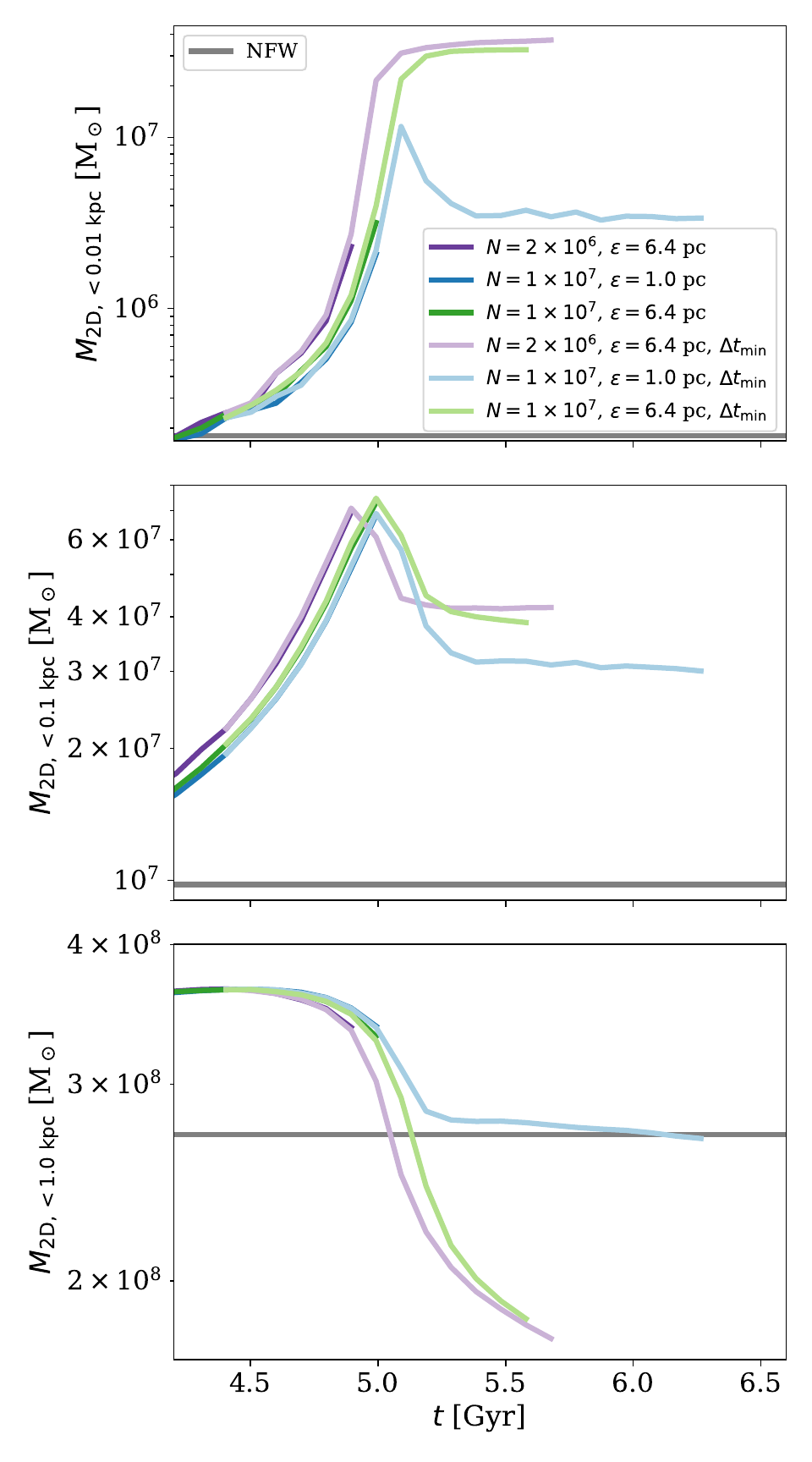}
    \caption{Enclosed mass within several radii as a function of time. The results are shown for simulations with the standard time step criterion (darker curves) and those employing a minimum time step (lighter curves). The three panels give the enclosed mass within different radii for the same simulations, as shown in Fig.~\ref{fig:min_tstep}. Moreover, we show the enclosed mass for the initial NFW halo as a reference (grey).
    }
    \label{fig:enclosed_mass_min_tstep}
\end{figure}

In addition to the results above, we also show how the enclosed mass evolves for our simulations employing a minimum time step. The results for all three of them are given in Fig.~\ref{fig:enclosed_mass_min_tstep}. The figure shows that for the late phase that we would hardly be able to simulate without imposing a minimum time step; the values for the enclosed masses differ significantly between the different runs.
As a result, cosmological simulations employing a minimum for the allowed time step may overestimate the collapse time and underestimate the density at the halo centre. But importantly, they may also overestimate the mass of the halo within the scale radius of the initial NFW profile or larger radii. As a consequence, the ability of SIDM simulations to explain gravitational lensing signals supposedly arising from fairly concentrated objects can be easily overestimated.

\subsection{Velocity and angular dependence}
\label{sec:angular_and_velocity_dep_iso}
\label{sec:}
\begin{figure}
    \centering
    \includegraphics[width=\columnwidth]{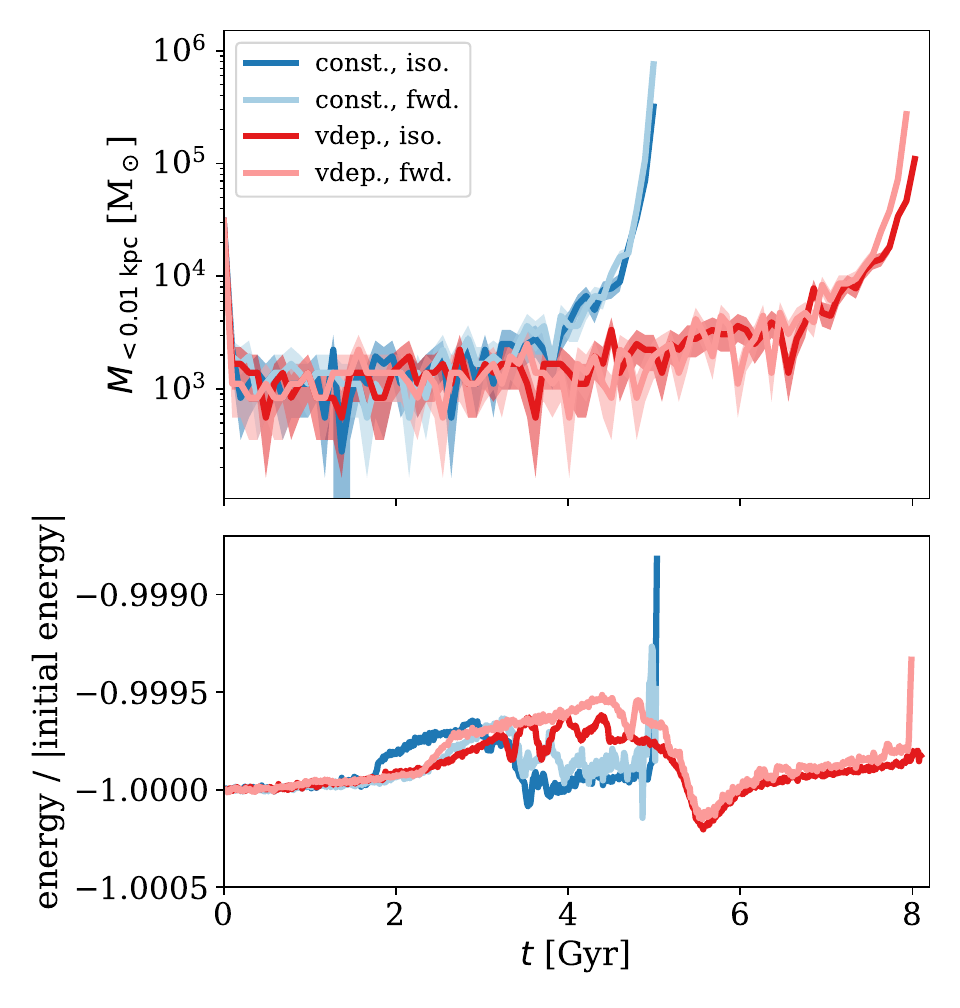}
    \caption{Velocity and angular dependent cross-sections. Following Fig.~\ref{fig:soft_tstep}, the enclosed mass within 10 pc (upper panel) and the energy conservation (lower panel) are shown. In detail, a velocity-independent isotropic cross-section (blue), a velocity-independent forward-dominated cross-section (light blue), a velocity-dependent isotropic cross-section (red), and a velocity-dependent forward-dominated cross-section (light red) are displayed. They all have the same resolution of $N=10^7$ and a gravitational softening length of $\epsilon = 1 \, \mathrm{pc}$. More details on the simulations O, P, R, and S can be found in Table~\ref{tab:sim_para}.}
    \label{fig:vdep+angle}
\end{figure}

So far, we have discussed the simulation results for an isotropic velocity-independent cross-section only. Now we focus on the angular and velocity dependence.
In contrast to the previous part, we are mainly interested in qualitative differences arising from particle physics. Given that all the simulations we discuss here are based on similar numerical schemes, they also have similar numerical properties.

We simulated cross-sections with different velocity and angular dependences. 
For the velocity-dependent cross-sections we used
\begin{equation} \label{eq:vel_dep}
    \frac{\sigma_\mathrm{V}}{m} = \frac{\sigma_0}{m} \left[ 1 + \left( \frac{v}{w} \right)^2 \right]^{-2} \, .
\end{equation}
Here, $\sigma_\mathrm{V}$ is the viscosity cross-section as given by Eq.~\eqref{eq:viscosity_cross_section}. 
The velocity-dependent cross-sections we simulated are given by $\sigma_\mathrm{0}/m = 6593.89 \, \mathrm{cm}^2\mathrm{g}^{-1}$ (in terms of the normalised viscosity cross-section, Eq.~\eqref{eq:viscosity_cross_section}) and $w=20 \, \mathrm{km} \, \mathrm{s}^{-1}$.
All values used for our simulations are given in Table~\ref{tab:sim_para}.
Moreover, we simulated two different angular dependences, isotropic scattering and a forward-dominated cross-section, specifically the limit of keeping the momentum transfer constant while the scattering angle approaches zero \citep[e.g.][]{Kahlhoefer_2014, Fischer_2021a}.

In Fig.~\ref{fig:vdep+angle}, we display the evolution of the isolated halo with different velocity and angular dependences. The angular dependences were matched using the viscosity cross-section (Eq.~\eqref{eq:viscosity_cross_section}). As we can see, this results in a similar time evolution for the velocity-independent and dependent cases. Moreover, the energy error is similar as well.
This demonstrates that the viscosity cross-section works very well for matching different angular dependences, as previously stated by \cite{Yang_2022D, Sabarish_2024}.
In contrast to the angular dependence, the velocity dependence (Eq.~\eqref{eq:vel_dep}) leads to a qualitatively different evolution of the halo, with a larger maximum core size and a longer collapse time compared to the core expansion phase. This qualitative behaviour arising from the velocity dependence is discussed in greater detail in \citet{Fischer_2024a} (Sect.~3.2.3). We provide further discussion in Appendix~\ref{sec:sigma_eff_iso}. 

\subsection{Velocity anisotropy} \label{sec:anisotropy}

\begin{figure}
    \centering
    \includegraphics[width=\columnwidth]{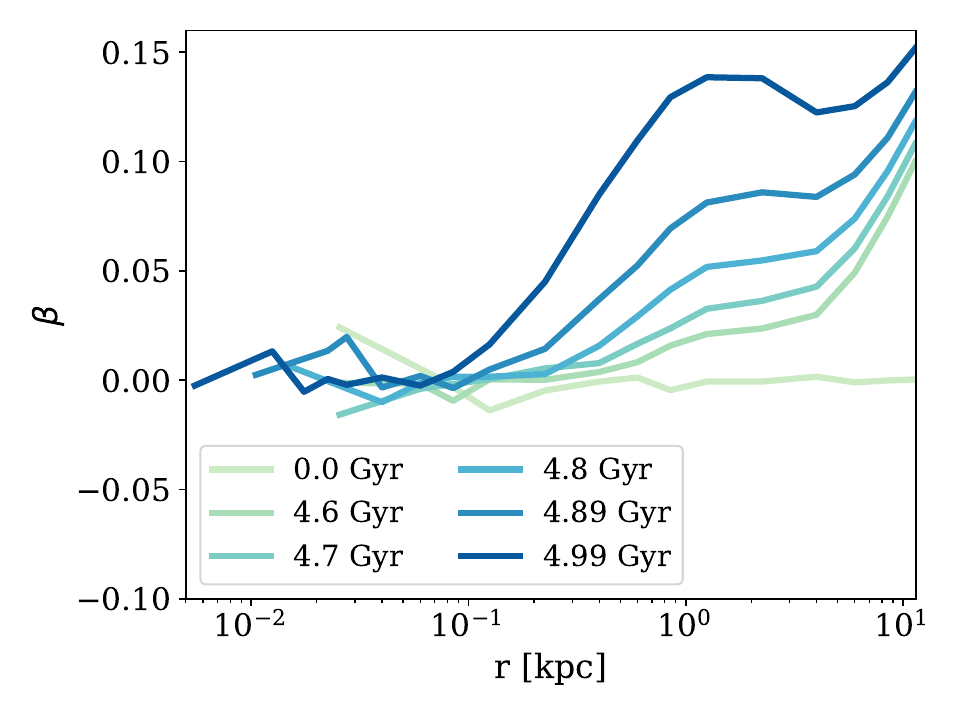}
    \caption{Velocity anisotropy as a function of radius. We show $\beta$ (Eq.~\eqref{eq:anisotropy}) for our most highly resolved simulation (simulation Q of Table~\ref{tab:sim_para}) at different times.}
    \label{fig:anisotropy}
\end{figure}

As a last part on the evolution of the isolated halo, we present the velocity anisotropy, and compute it as
\begin{equation} \label{eq:anisotropy}
    \beta \equiv 1 - \frac{\sigma^2_\theta + \sigma^2_\phi}{2 \, \sigma^2_r} \,,
\end{equation}
with the polar velocity dispersion $\sigma^2_\theta$, the azimuthal velocity dispersion $\sigma^2_\phi$, and the radial velocity dispersion $\sigma^2_r$.
The results are displayed in Fig.~\ref{fig:anisotropy}.
For $t=0$ the velocity distribution is isotropic at all radii ($\beta = 0$), but increases at later times at the larger radii beyond the inner density core. Here, the radial motion becomes more pronounced (see also \cite{Gurian_2025}).
We note that at the largest radii shown in Fig.~\ref{fig:anisotropy}, the anisotropy increases. This might be a consequence of the particles moving radially outwards becoming more dominant as the density decreases strongly with radius ($\rho \propto r^{-3}$). In addition, the collision rate is low at those radii, and the self-interactions do not efficiently isotropise the velocity distribution.

\section{Satellite evolution} \label{sec:results2}

In this section we discuss the evolution of the SIDM halo under the influence of tidal forces arising from a host galaxy. We first describe the changes to our numerical set-up in Sect.~\ref{sec:satellite_setup}.
An illustration of the satellite orbit and the tidal evolution can be found in Appendix~\ref{sec:orbit_and_tidal_evolution}.
We show the difference between a satellite halo and an isolated system (Sect.~\ref{sec:isolated_vs_satellite}) as well as the difference arising from the angular and velocity dependence (Sect.~\ref{sec:angular_and_velocity_dep_sat}).
Moreover, we demonstrate in Sect.~\ref{sec:king_fit_sat} that the density profile during the collapse phase can be well described by a King model.
Overall, we mainly focus on the physics, in contrast to the numerical aspects of the previous section. Given that we tested various numerical parameters for the isolated case, we now assume that the same choice also works when evolving the halo in an external potential. 
We note that we used the peak find algorithm by \cite{Fischer_2021b} to determine the position of the satellite. As in the previous section, we compute physical quantities with respect to this position for all of this section.

\subsection{Numerical set-up} \label{sec:satellite_setup}

To model the tidal forces of the host system acting on the satellite at low computational costs, we described the gravitational potential of the host analytically. We focused only on the role of the tidal forces and neglected the dynamical friction that acts on the satellite and causes its orbit to decay. Moreover, we also did not take the scattering of DM particles of the satellite halo on the host's DM into account, which can play an important role especially for velocity-independent cross-sections \citep[e.g.][]{Zeng_2022}.

The external potential is chosen to mimic the gravitational potential of the Milky Way and consists of six components.
The details of our description follow that employed by \cite{Zhang_2025}.
We describe the DM halo with an NFW profile \citep{Navarro_1996}. The corresponding potential is
\begin{equation} \label{eq:nfw_potential}
    \Phi_\mathrm{NFW}(r) = -4 \uppi \, \mathrm{G} \, \frac{\rho_\mathrm{s} \, r_\mathrm{s}^3}{r} \, \ln\left(1 + \frac{r}{r_\mathrm{s}}\right) \,.
\end{equation}
Here G denotes the gravitational constant.
For our simulations, we set the NFW halo parameters to $\rho_s = 8.54 \times 10^{6} \, \mathrm{M_\odot} \, \mathrm{kpc}^{-3}$ and $r_\mathrm{s} = 19.6 \, \mathrm{kpc}$.
To mimic the gravitational potential of the stellar bulge, we employed a Hernquist profile \citep{Hernquist_1990}.
\begin{equation} \label{eq:hernquist_potential}
    \Phi_\mathrm{Hern}(r) = - \frac{\mathrm{G} M_\mathrm{H}}{a_\mathrm{H}} \, \left(1 + \frac{r}{a_\mathrm{H}}\right)^{-1} \, .
\end{equation}
For our set-up, we used $M_\mathrm{H} = 9.23 \times 10^9 \, \mathrm{M_\odot}$ and $a_\mathrm{H} = 1.3 \, \mathrm{kpc}$.
In addition, we added four disk components described by an axisymmetric Miyamoto–Nagai profile \citep{Miyamoto_1975},
\begin{equation} \label{eq:disk_potential}
    \Phi_\mathrm{disk}(r) = - \mathrm{G} \, M_\mathrm{d} \left[ R^2 + \left( a_\mathrm{d} + \sqrt{z^2 + b^2_\mathrm{d}} \right)^2 \right]^{-1/2},
\end{equation}
with $R = \sqrt{x^2 + y^2}$. The components are as follows:
a thin stellar disk with $M_\mathrm{d} = 3.52 \times 10^{10} \,\mathrm{M_\odot}$, $a_\mathrm{d} = 2.5 \,  \mathrm{kpc}$,
and $b_\mathrm{d} = 0.3 \, \mathrm{kpc}$;
a thick stellar disk with $M_\mathrm{d} = 1.05 \times 10^{10} \, \mathrm{M_\odot}$, $a_\mathrm{d} = 3.02 \, \mathrm{kpc}$, and $b_\mathrm{d} = 0.9 \, \mathrm{kpc}$;
a thin gas disk with $M_\mathrm{d} = 1.2 \times 10^9 \,\mathrm{M_\odot}$, $a_\mathrm{d} = 1.5 \, \mathrm{kpc}$,
and $b_\mathrm{d} = 0.045 \, \mathrm{kpc}$;
and last a thick gas disk with $M_\mathrm{d} = 1.1 \times 10^{10} \, \mathrm{M_\odot}$, $a_\mathrm{d} = 7.0 \, \mathrm{kpc}$, and $b_\mathrm{d} = 0.085 \, \mathrm{kpc}$.

The satellite halo is initially placed at a distance of $r=123.6 \, \mathrm{kpc}$ from the centre of the external potential. The exact coordinates are $[58.16, 40.12, 101.41] \, (\mathrm{kpc})$. The position is relative to the origin of the external potential. The initial velocity of the satellite is $v = 109 \, \mathrm{km} \, \mathrm{s}^{-1}$. The velocity vector reads $[45.94, 83.42, 53.12] \, (\mathrm{km} \, \mathrm{s}^{-1})$. With this choice, we also follow \cite{Zhang_2025}. Moreover, we illustrate the orbit of the satellite halo in Appendix~\ref{sec:orbit_and_tidal_evolution}.

\subsection{Isolated versus satellite}
\label{sec:isolated_vs_satellite}

\begin{figure}
    \centering
    \includegraphics[width=\columnwidth]{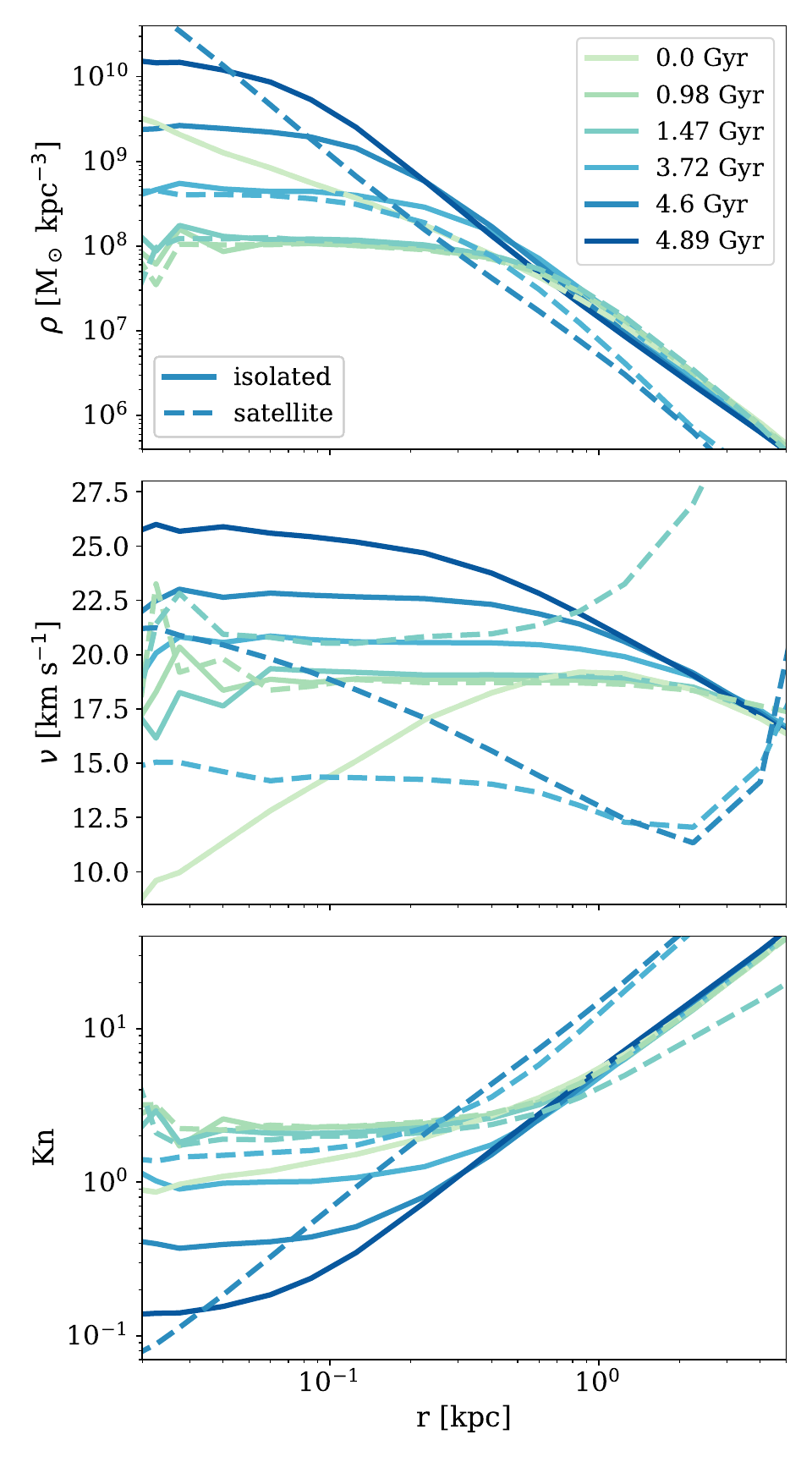}
    \caption{Evolution of an isolated halo (solid line) and a satellite halo (dashed line). The simulations are for a velocity-independent cross-section, employ $N=10^7$ particles, and use a gravitational softening length of $\epsilon=1\,\mathrm{pc}$. All parameters for the displayed simulations O and W are given in Table~\ref{tab:sim_para}.
    }
    \label{fig:profiles_iso_vs_sat}
\end{figure}

Next, we discuss how the evolution of the density and velocity dispersion profiles changes for a satellite halo compared to a system evolved in isolation. 
In Fig.~\ref{fig:profiles_iso_vs_sat}, we show the evolution of the density, the one-dimensional velocity dispersion, and the Knudsen number, as a function of radius for a halo in isolation (solid lines) and for a satellite system (dashed lines).

By and large, the evolution of the satellite system is similar to that of the isolated halo. First, the two undergo a core expansion phase and later collapse. However, the velocity dispersion profile (middle panels of Fig.~\ref{fig:profiles_iso_vs_sat}) reveals a significant difference. For the satellite halo, the velocity dispersion is strongly affected by tidal forces. Close to the pericentre passage (e.g.\ at $1.47\,\mathrm{Gyr}$) tidal heating injects energy into the halo and the velocity dispersion increases. Subsequently, particles become unbound and leave the system, and the velocity dispersion decreases over all relevant radii. This is in contrast to the isolated halo, where the velocity dispersion is only affected by the self-interactions.
The speed-up of the halo evolution due to tidal stripping has been found by various authors \citep[e.g.][]{Kahlhoefer_2019, Sameie_2020b}.
This makes the energy outflow of the satellite more efficient and speeds up the halo evolution, i.e.\ reduces the collapse time.
Later, in the collapse phase, the velocity dispersion in the centre reaches unprecedentedly high values.

We note that the satellite system reaches a central density of $\rho = 3 \times 10^{11} \, \mathrm{M_\odot} \, \mathrm{kpc}^{-3}$ in the last snapshot at $t = 4.6 \, \mathrm{Gyr}$.
At sufficiently small radii, the density gradient becomes flat and the Knudsen number reaches a value of about 0.04, i.e.\ the central region is deep in the short-mean-free-path regime.
In contrast, the lowest central density we find at the maximum core expansion is about $\rho = 6 \times 10^7 \, \mathrm{M_\odot} \, \mathrm{kpc}^{-3}$. It is lower compared to the isolated case due to tidal interactions. Moreover, the tidal interactions also greatly reduce the density in the outer regions of the halo, which leads to a large loss in gravitationally bound mass.

\subsection{Angular and velocity dependence of the cross-section} \label{sec:angular_and_velocity_dep_sat}

\begin{figure}
    \centering
    \includegraphics[width=\columnwidth]{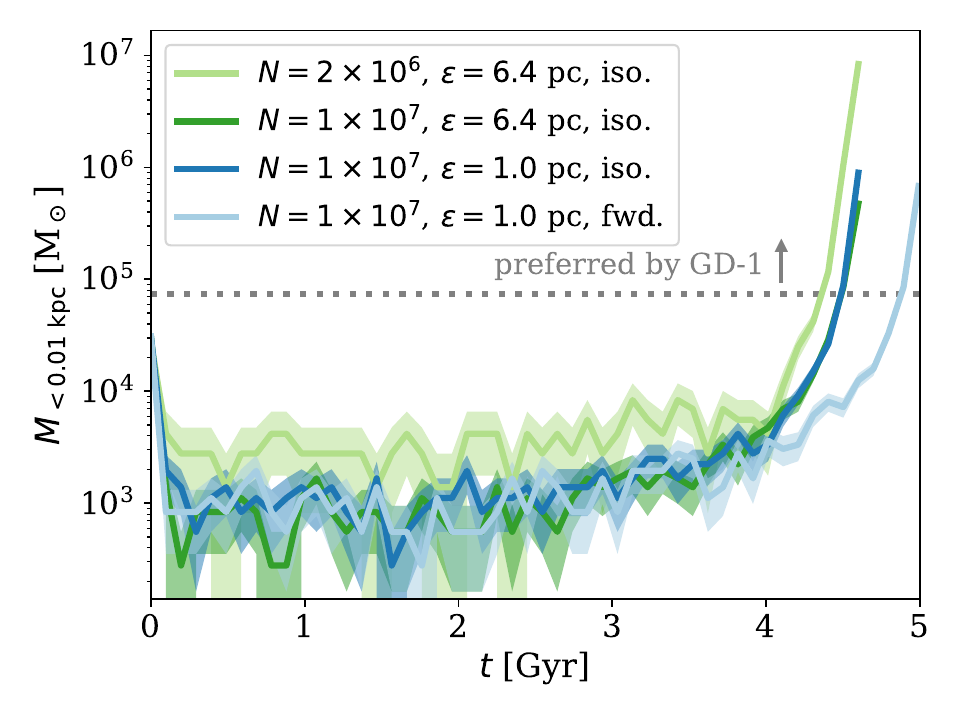}
    \includegraphics[width=\columnwidth]{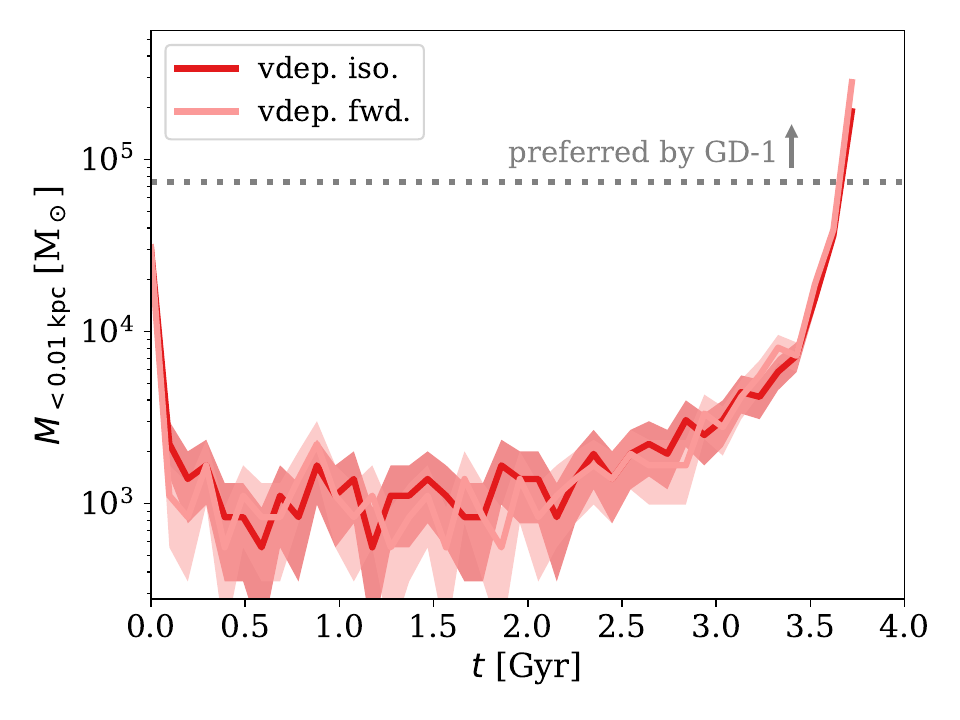}
    \caption{Mass enclosed within 10 pc as a function of time for the satellite halos. The upper panel gives the results for a velocity-independent cross-section (simulations U, V, W, and X), and the lower panel is for velocity-dependent cross-sections (simulations Y and Z). More details on the simulations can be found in Table~\ref{tab:sim_para}. In addition, the grey dotted line indicates the preferred mass range for the perturber of the stellar stream GD-1 \citep{Bonaca_2019}.}
    \label{fig:sat_mass_enclosed}
\end{figure}

In this section we investigate the role of the velocity and angular dependence of the self-interaction for the evolution of the satellite halo. As for the isolated halo, we employed isotropic and forward-dominated cross-sections that either follow the velocity dependence given by Eq.~\eqref{eq:vel_dep} or are velocity-independent. In detail, we employed the same cross-section as in Sect.~\ref{sec:angular_and_velocity_dep_iso}.
The results for the enclosed mass within 10 pc are shown in Fig.~\ref{fig:sat_mass_enclosed}.

The tidal forces acting on the satellite halo only slightly speed up the gravothermal evolution in the case of a velocity-independent cross-section.
Interestingly, we find that this is very different for our velocity-dependent cross-sections. When the halo experiences tidal forces, the collapse time reduces to about half its value compared to the evolution in isolation (compare Figs.~\ref{fig:vdep+angle} and~\ref{fig:sat_mass_enclosed}).

This implies that a velocity-dependent cross-section not only speeds up the collapse by suppressing interactions between DM particles of the host and the satellite, but there is already a contribution coming from within the satellite halo. We note that our cross-section has a fairly strong velocity dependence, i.e.\ $w$ is smaller than the relative velocities within the halo. This implies an increase in the effective cross-section relative to the velocity-independent case when the velocity dispersion decreases due to tidal stripping.
We show this in detail in Appendix~\ref{sec:sigma_eff_sat}.

Interestingly, we find that the angular dependence starts to play a role in the velocity-independent case, although the scatterings between the DM of the host and the satellite are neglected. The small-angle scattering leads here to a longer collapse time compared to the isotropic cross-section (Fig.~\ref{fig:sat_mass_enclosed}). For the isolated halo, the two cross-sections resulted in the same collapse time (Fig.~\ref{fig:vdep+angle}). \cite{Fischer_2021a} show that the thermalisation process depends on the angular dependence of the cross-section, i.e.\ large-angle scattering populates the high-velocity tail of the Maxwell-Boltzmann distribution more quickly compared to small-angle scattering. In combination with the tidal forces, this allows us to explain the different collapse times.
Moreover, the difference arising from the angular dependence is minor in the velocity-dependent case where scattering with a high relative velocity is strongly suppressed (bottom panel of Fig.~\ref{fig:sat_mass_enclosed}). We also checked that both angular-dependent and -independent cases give rise to almost the same total gravitationally bound mass of the satellite halo regardless of the velocity dependence. For the runs shown here, the angular dependence of the scattering does not affect the global properties of the halo. 

\subsection{Fit with King model}
\label{sec:king_fit_sat}
\begin{figure}
    \centering
    \includegraphics[width=\columnwidth]{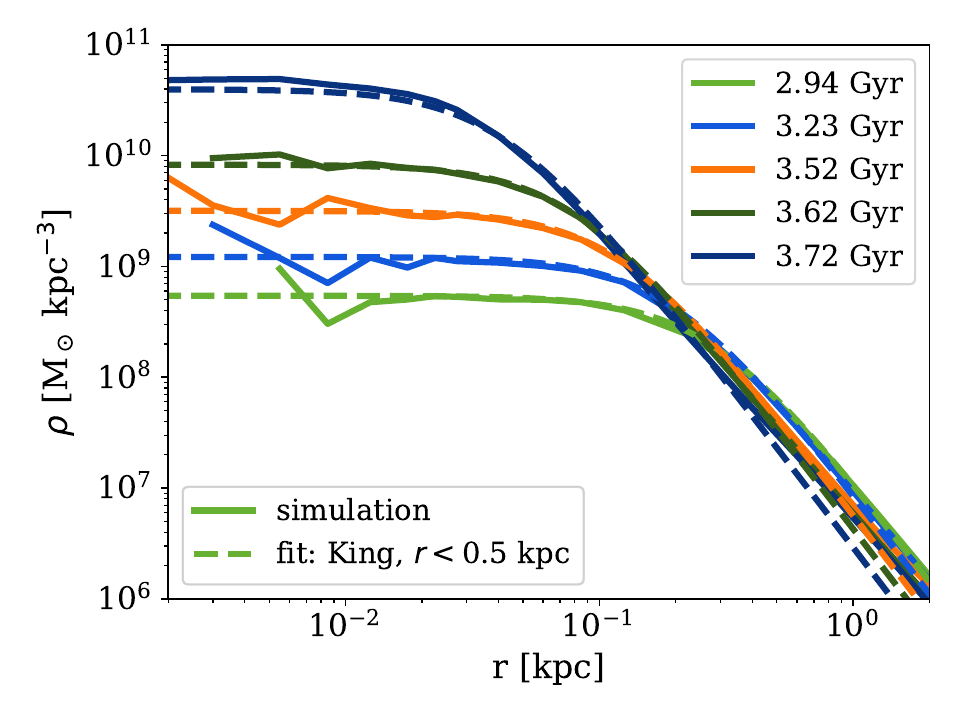}
    \caption{Density profile fitted by a King model for a satellite halo. The results are for the simulation with a velocity-dependent isotropic cross-section. All simulation parameters are given in Table~\ref{tab:sim_para} under simulation Y.}
    \label{fig:fit_king_sat}
\end{figure}
In this last part about the satellite halo simulations, we fit the density profile of the late collapse phase with a King model \citep{King_1962},
\begin{equation} \label{eq:king}
    \rho(r) = \rho_0 \left[ 1 + \left( \frac{r}{r_\mathrm{c}} \right)^2 \right]^{-3/2} \,.
\end{equation}
With this, we follow \cite{Zhang_2025} who also used a King profile to describe the collapse phase of a potential GD-1 perturber, motivated by research on star clusters.
To determine the density and radial parameters ($\rho_0$, $r_\mathrm{c}$), we used a limited radial range with $r<0.5 \, \mathrm{kpc}$ only.
For the fit, we maximised a likelihood based on Poisson statistics analogously to the description in Sect.~4 by \cite{Fischer_2021a}.

In Fig.~\ref{fig:fit_king_sat}, we show the fit of the King model for the collapse phase using our simulation with a velocity-dependent isotropic cross-section (simulation Y according to Table~\ref{tab:sim_para}). It becomes visible that the King model provides a reasonable fit to the inner region of a satellite halo that undergoes core collapse.
This is in line with the findings of \cite{Zhang_2025} and may work well enough to fit the observational data, as discussed by the same authors.
In addition, we fit the King mode to the isolated halo and show the density profile as well as the time evolution of the fitted parameter
in Appendix~\ref{sec:king_fit}.

\section{Discussion} \label{sec:discussion}

In this section we discuss the limitations of our work and of the employed numerical scheme. We mention the physical processes that we neglected and that would be worth including in the modelling, and we highlight directions to improve the modelling of collapsing SIDM halos.

In our simulations we neglected the scatterings between satellite and host particles. However, they can play a crucial role in shaping the evolution of the satellite. How strongly the DM of the satellite is affected depends on how strong the cross-section is on the relevant velocity scale, which is typically larger than for the velocities within the satellite. Hence, for a cross-section that is decreasing as a function of velocity, this is less important than for a velocity-independent cross-section \citep[e.g.][]{Silverman_2022, Zeng_2022, Zeng_2025a}. In addition, the angular dependence of the cross-section can play an important role. Satellite halos are strongly affected if the cross-section is forward-dominated \citep{Fischer_2022} and the effect becomes weaker for velocity-dependent cross-sections \citep[e.g.][]{Nadler_2020, Fischer_2024a}.
The DM self-interactions between host and satellite could be incorporated in simulations such as the ones we have presented following the approach by \cite{Zeng_2022} or within the more general framework of Klemmer et al.\ (in prep.).

Another physical effect that we neglect here is dynamical friction. However, given the low mass of our satellite halo compared to the host mass, it is expected to be hardly affected by dynamical friction. Nevertheless, it would be possible to include the effect of dynamical friction without resolving the host system, but by approximating the deceleration experienced by the subhalo \citep[e.g.][]{Petts_2015}.

Our numerical scheme has the advantage that the pairwise interactions do not harm energy conservation, although we employ shared and distributed memory parallelisation. Many of the codes used today do not have this property.
In our case, it implies that the energy conservation is independent of the chosen kernel function for the SIDM computations, by construction. This is because the value of the interaction probability or the drag force does not have a direct impact on energy conservation. However, the chosen kernel function could eventually make a difference in the context of angular momentum conservation and for the effective heat conduction in the case of large values for $h/l$ (see Sect.~\ref{sec:kernel_size}).

Ways to improve the accuracy of SIDM simulations have already been discussed by \cite{Fischer_2024a}. We want to highlight that using a symmetric evaluation of the oct-tree used for the gravitational force computation would be helpful (e.g.\ as done by \cite{Appel_1985}).
Another aspect is the time-irreversible choice of the time step. Approaches that make the time-stepping function more time symmetric can help to improve the accuracy, for example by reducing the error on energy conservation \citep{Dehnen_2017}.

We also note that the above-mentioned strategies may not fully solve the challenges in simulating the late collapse phase. For example, it remains problematic to resolve smaller and smaller length scales while the halo is collapsing. Even if this were resolved, and all numerical issues directly related to simulating SIDM as well, the modelling of gravity would still require a time step that continues shrinking while the halo is collapsing. The time step would become so small that the simulation could no longer be reasonably fast advanced in time.
One way to overcome the challenges posed by gravothermal collapse could be the introduction of a subgrid model that describes the very inner part of the halo. It might be a sink particle that accounts for the DM mass in the centre of the halo. In the post-collapse state, it may describe the black hole that finally formed and the DM spike that it is accreting. Idealised simulations of such a system can help build such a model \citep{Sabarish_2025}.

\section{Conclusion} \label{sec:conclusion}

In this work, we used the $N$-body simulation code \textsc{OpenGadget3} to investigate the gravothermal collapse of SIDM halos.
We studied the role of various numerical parameters and simulated the system in isolation and as a satellite halo. Additionally, we studied the qualitative differences arising from the velocity and angular dependence of the self-interaction cross-section. 
Furthermore, the data for our most highly resolved simulation ($N = 5 \times 10^7$) are available\footnote{The simulation data can be found at \url{https://darkium.org}.} and may serve as a benchmark.
Our main results and recommendations for SIDM simulations are as follows:

\begin{enumerate}
    \item A very small softening length can lead to larger errors in the total energy because the time steps change more often in an asymmetric way. However, the softening length chosen should be  small enough to resolve the relevant length scales (e.g.\ the size of the constant density core during collapse).
    \item Overall, halo evolution is very sensitive to energy conservation errors.
    \item We did not find that a stricter gravity time step criterion is needed for our SIDM simulations compared to what is commonly used for CDM simulations, i.e.\ $\eta = 0.025$ is fine. At the same time, we may use a more stringent SIDM time step criterion than some other codes. It should be noted that simulations that evolve an SIDM halo over many more gravitational relaxation timescales than we did may need a smaller value for $\eta$ \cite[see also Sect.~4 by][]{Fischer_2024b}.
    \item Moreover, we recommend using a value for the opening criterion as small as $\alpha = 5\times 10^{-4}$, when using a one-sided oct-tree.
    \item A SIDM kernel size that is too large effectively leads to heat conduction that is too strong, which causes the halo to collapse faster. We find that $h/l$ should be below 10, but exceeding 1 by a few multiples appears to be unproblematic for the range that we have tested. This has implications on the number of particles required to resolve the collapse phase, i.e.\ it depends on how deep one wants to simulate in the gravothermal collapse.
    \item We also note that the maximum sampling radius of ICs has an influence on how fast the halo collapses. For a controlled simulation of an isolated NFW halo, we recommend using 15 $r_s$ instead of 10 $r_s$ for the truncation radius.
    \item In our tests that enforce a minimum time step, we found that the central density increases further and stays constant at the same time that the total energy artificially increases.
    Although imposing a minimum time step in cosmological simulations may allow correct predictions of the fraction of collapsed halos to be made, we do not find evidence that the enclosed mass within a specific radius can be accurately predicted.
    The mass within larger radii ($\gtrsim r_\mathrm{s}$) might be overestimated.
    \item During the late collapse stages the velocity distribution is no longer isotropic, but becomes progressively more radial outside the inner density core.
    \item The speed-up in halo evolution for a satellite compared to an isolated system is much greater for a velocity-dependent cross-section compared to a velocity-independent one, even when ignoring the scattering between the DM particles of the satellite with the host's DM.
    \item We find that a King model provides a good fit for the density profile in the inner regions of the halo undergoing gravothermal collapse at the stage that we have simulated. This is the case for the isolated halo and the satellite.
\end{enumerate}

Despite the numerical challenges, we have shown that it is possible to accurately simulate SIDM halos in the deep-collapse phase to compare with observations. Taking the dense perturber of the GD-1 stellar stream as an example, our simulations can resolve the simulated halo within 10 pc, which is necessary to match the observations. It becomes prohibitively expensive to simulate extremely deep into the collapse phase using the $N$-body method, such as collapsing into a black hole. Nevertheless, in that regime, simulations based on schemes derived from first principles exploiting the symmetries of the problem can provide a complementary method \citep[e.g.][]{Gurian_2025, Kamionkowski_2025}.

\begin{acknowledgements}
We thank all participants of the Darkium SIDM Journal Club for discussion, in particular Antonio Ragagnin. MSF is also grateful to Andrew Robertson for his comments and suggestions.
This work is funded by the Deutsche Forschungsgemeinschaft (DFG, German Research Foundation) under Germany’s Excellence Strategy -- EXC-2094 ``Origins'' -- 390783311.
KD and MSF acknowledge support by the COMPLEX project from the European Research Council (ERC) under the European Union’s Horizon 2020 research and innovation program grant agreement ERC-2019-AdG 882679.
HBY acknowledges support by the U.S.\ Department of Energy under grant No.\ de-sc0008541.

Software:
NumPy \citep{NumPy},
Matplotlib \citep{Matplotlib}.
\end{acknowledgements}

\bibliographystyle{aa}
\bibliography{bib}

\begin{appendix}
\section{Overview of simulations}
\label{sec:sim_overview}
\begin{table*}
    \caption{Simulation properties and parameters.}
    \label{tab:sim_para}
    \centering
    \begin{tabular}{cccccccccccc}
        \hline\hline
        Sim. & $\sigma_\mathrm{V} / m$ & $w$ & angular& $m_\mathrm{n}$ & $N$ & $r_\mathrm{cut}$ & $\epsilon$ & $\eta$ & $N_\mathrm{ngb}$ & $\tau$ & $\Delta t_\mathrm{min}$\\
        name & $[\mathrm{cm}^2 \mathrm{g}^{-1}]$ & [km s$^{-1}$] & depend. & [M$_\odot$] & & [$r_\mathrm{s}$] & [pc] & & & & [Gyr]\\ \hline
        {\color{simA} \textbf{A}} & 0 & -- & -- & $1.39 \times 10^{3}$ & $2 \times 10^6$ & 15 & 6.4 & $2.5 \times 10^{-2}$ & -- & -- & -- \\
        {\color{simB} \textbf{B}} & 0 & -- & -- & $1.39 \times 10^{3}$ & $2 \times 10^6$ & 15 & 1.0 & $2.5 \times 10^{-2}$ & -- & -- & -- \\
        {\color{simC} \textbf{C}} & 0 & -- & -- & $1.39 \times 10^{3}$ & $2 \times 10^6$ & 15 & 1.0 & -- & -- & -- & $2.2 \times 10^{-5}$ \\
        {\color{simD} \textbf{D}} & 0 & -- & -- & $2.78 \times 10^{2}$ & $1 \times 10^7$ & 15 & 6.4 & $2.5 \times 10^{-2}$ & -- & -- & -- \\
        {\color{simE} \textbf{E}} & 0 & -- & -- & $2.78 \times 10^{2}$ & $1 \times 10^7$ & 15 & 1.0 & $2.5 \times 10^{-2}$ & -- & -- & -- \\
        {\color{simF} \textbf{F}} & 0 & -- & -- & $5.57 \times 10^{1}$ & $5 \times 10^7$ & 15 & 1.0 & $2.5 \times 10^{-2}$ & -- & -- & -- \\
        {\color{simG} \textbf{G}} & 80 & -- & iso. & $1.39 \times 10^{3}$ & $1.69 \times 10^6$ & 10 & 6.4 & $2.5 \times 10^{-2}$ & 48 & 0.04 & -- \\
        {\color{simH} \textbf{H}} & 80 & -- & iso. & $1.39 \times 10^{3}$ & $2 \times 10^6$ & 15 & 6.4 & $1 \times 10^{-3}$ & 48 & 0.04 & -- \\
        {\color{simI} \textbf{I}} & 80 & -- & iso. & $1.39 \times 10^{3}$ & $2 \times 10^6$ & 15 & 6.4 & $5 \times 10^{-3}$ & 48 & 0.04 & -- \\
        {\color{simJ} \textbf{J}} & 80 & -- & iso. & $1.39 \times 10^{3}$ & $2 \times 10^6$ & 15 & 6.4 & $2.5 \times 10^{-2}$ & 48 & 0.04 & -- \\
        {\color{simK} \textbf{K}} & 80 & -- & iso. & $1.39 \times 10^{3}$ & $2 \times 10^6$ & 15 & 6.4 & $2.5 \times 10^{-2}$ & 384 & 0.04 & -- \\
        {\color{simL} \textbf{L}} & 80 & -- & iso. & $1.39 \times 10^{3}$ & $2 \times 10^6$ & 15 & 1.0 & $2.5 \times 10^{-2}$ & 48 & 0.04 & -- \\
        {\color{simM} \textbf{M}} & 80 & -- & iso. & $2.78 \times 10^{2}$ & $8.45 \times 10^6$ & 10 & 1.0 & $2.5 \times 10^{-2}$ & 48 & 0.04 & -- \\
        {\color{simN} \textbf{N}} & 80 & -- & iso. & $2.78 \times 10^{2}$ & $1 \times 10^7$ & 15 & 6.4 & $2.5 \times 10^{-2}$ & 48 & 0.04 & -- \\
        {\color{simO} \textbf{O}} & 80 & -- & iso. & $2.78 \times 10^{2}$ & $1 \times 10^7$ & 15 & 1.0 & $2.5 \times 10^{-2}$ & 48 & 0.04 & -- \\
        {\color{simP} \textbf{P}} & 80 & -- & fwd. & $2.78 \times 10^{2}$ & $1 \times 10^7$ & 15 & 1.0 & $2.5 \times 10^{-2}$ & 48 & 0.04 & -- \\
        {\color{simQ} \textbf{Q}} & 80 & -- & iso. & $5.57 \times 10^{1}$ & $5 \times 10^7$ & 15 & 1.0 & $2.5 \times 10^{-2}$ & 48 & 0.04 & -- \\
        {\color{simJt} \textbf{Jt}} & 80 & -- & iso. & $1.39 \times 10^{3}$ & $2 \times 10^6$ & 15 & 6.4 & $2.5 \times 10^{-2}$ & 48 & 0.04 & $2.8 \times 10^{-5}$ \\
        {\color{simNt} \textbf{Nt}} & 80 & -- & iso. & $2.78 \times 10^{2}$ & $1 \times 10^7$ & 15 & 6.4 & $2.5 \times 10^{-2}$ & 48 & 0.04 & $2.8 \times 10^{-5}$ \\
        {\color{simOt} \textbf{Ot}} & 80 & -- & iso. & $2.78 \times 10^{2}$ & $1 \times 10^7$ & 15 & 1.0 & $2.5 \times 10^{-2}$ & 48 & 0.04 & $2.8 \times 10^{-5}$ \\
        {\color{simR} \textbf{R}} & 6593.89 & 20 & iso. & $2.78 \times 10^{2}$ & $1 \times 10^7$ & 15 & 1.0 & $2.5 \times 10^{-2}$ & 48 & 0.04 & -- \\
        {\color{simS} \textbf{S}} & 6593.89 & 20 & fwd. & $2.78 \times 10^{2}$ & $1 \times 10^7$ & 15 & 1.0 & $2.5 \times 10^{-2}$ & 48 & 0.04 & -- \\
        \hdashline
        {\color{simT} \textbf{T}} & 0 & -- & -- & $2.78 \times 10^{2}$ & $1 \times 10^7$ & 15 & 1.0 & $2.5 \times 10^{-2}$ & -- & -- & -- \\
        {\color{simU} \textbf{U}} & 80 & -- & iso. & $1.39 \times 10^{3}$ & $2 \times 10^6$ & 15 & 6.4 & $2.5 \times 10^{-2}$ & 48 & 0.04 & -- \\
        {\color{simV} \textbf{V}} & 80 & -- & iso. & $2.78 \times 10^{2}$ & $1 \times 10^7$ & 15 & 6.4 & $2.5 \times 10^{-2}$ & 48 & 0.04 & -- \\
        {\color{simW} \textbf{W}} & 80 & -- & iso. & $2.78 \times 10^{2}$ & $1 \times 10^7$ & 15 & 1.0 & $2.5 \times 10^{-2}$ & 48 & 0.04 & -- \\
        {\color{simX} \textbf{X}} & 80 & -- & fwd. & $2.78 \times 10^{2}$ & $1 \times 10^7$ & 15 & 1.0 & $2.5 \times 10^{-2}$ & 48 & 0.04 & -- \\
        {\color{simY} \textbf{Y}} & 6593.89 & 20 & iso. & $2.78 \times 10^{2}$ & $1 \times 10^7$ & 15 & 1.0 & $2.5 \times 10^{-2}$ & 48 & 0.04 & -- \\
        {\color{simZ} \textbf{Z}} & 6593.89 & 20 & fwd. & $2.78 \times 10^{2}$ & $1 \times 10^7$ & 15 & 1.0 & $2.5 \times 10^{-2}$ & 48 & 0.04 & -- \\
        \hline
    \end{tabular}
    \tablefoot{The table gives the different simulation properties and parameters that we employ. The first column gives the simulation name with the colour as used in the figures. It follows the second column with the viscosity cross-section Eq.~\eqref{eq:viscosity_cross_section}. The third column gives the velocity parameter of the velocity-dependent cross-section if applicable, in this case the second column corresponds to the normalisation of the cross-section, i.e.\ $\sigma_0$ of Eq.~\eqref{eq:vel_dep}. The fourth column gives the angular dependence (iso.\ -- isotropic, fwd.\ -- forward dominated). The fifth column specifies whether the halo is evolved in an external potential, and the following columns give the numerical particle mass, the number of particles, the cut radius for sampling the ICs, the gravitational softening length ($\epsilon$), the gravitational time step parameter ($\eta$), the SIDM neighbour number $N_\mathrm{ngb}$, the SIDM time step parameter ($\tau$), and the lower limit of the allowed time step ($\Delta t_\mathrm{min}$). Moreover, the parameter for the tree node opening criterion is set to $\alpha = 5 \times 10^{-4}$ for all simulations following \cite{Fischer_2024b}. The simulations above the dashed line are evolved in isolation, and the ones below in an external potential motivated by the Milky Way as described in Sect.~\ref{sec:satellite_setup}.
    The data for the highest-resolved SIDM simulation ($N=5\times10^7$) is available at \url{https://darkium.org}.}
\end{table*}
\FloatBarrier
With Table~\ref{tab:sim_para}, we provide an overview of the simulations shown in this paper. Moreover, we use a consistent colour scheme for the figures with the colours as indicated in the table.
The strength of the self-interactions is specified in terms of the viscosity cross-section. It is given by
\begin{equation} \label{eq:viscosity_cross_section}
\sigma_\mathrm{V} = 3 \uppi \int_{-1}^{1} \frac{\mathrm{d} \sigma}{\mathrm{d} \Omega_{\mathrm{cms}}}\sin^2\theta_{\mathrm{cms}} \, \mathrm{d} \cos \theta_{\mathrm{cms}}
\, ,
\end{equation}
and normalised to match the total cross-section in the case of isotropic scattering.

\section{The size of the time step}
\label{sec:grav_time_step}
\begin{figure}
    \centering
    \includegraphics[width=\columnwidth]{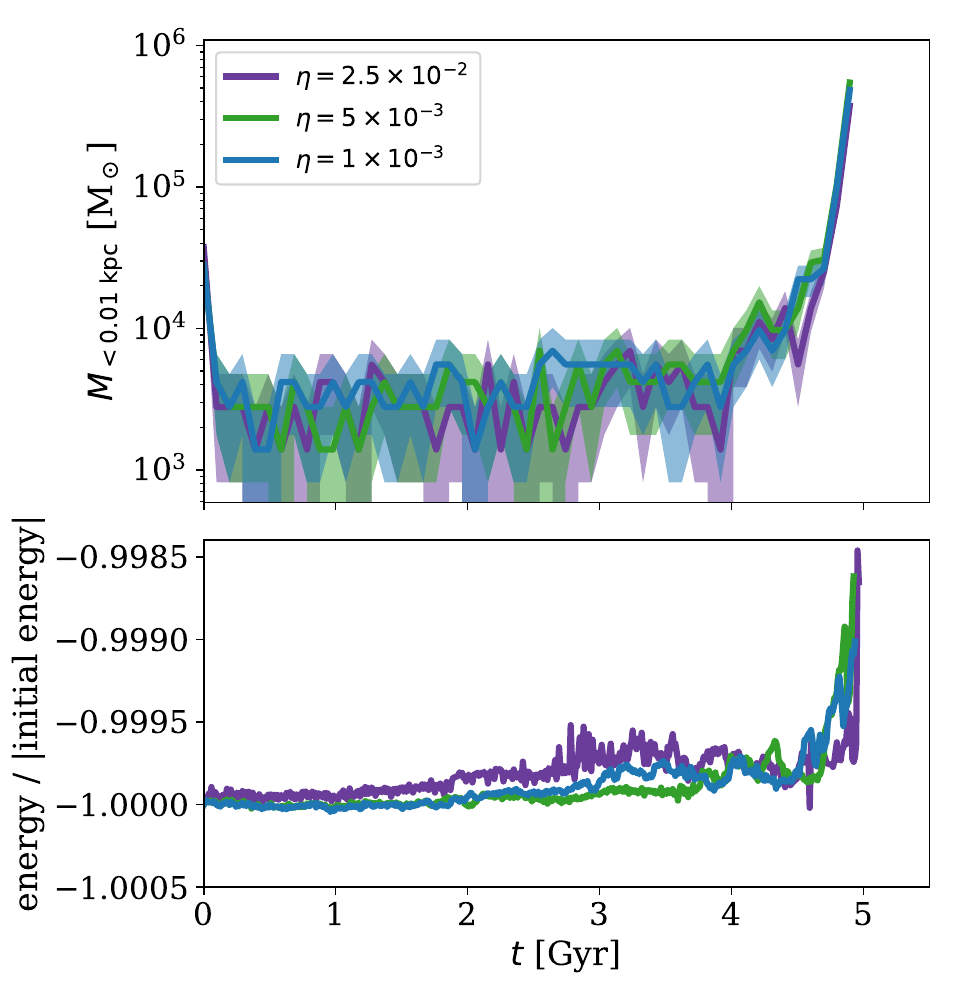}
    \caption{Variation of the gravitational time step parameter $\eta$. In the upper panel, we show the mass enclosed within the inner 10 pc as a function of time. The lower panel gives the evolution of the total energy relative to the absolute initial value. The shown simulations are for the low resolution of $N=2\times 10^6$ and share the same parameters except $\eta$, which is varied as indicated in the legend. All parameters are given in Table~\ref{tab:sim_para} under the names H, I, and J. The shaded regions in the top panel indicate the uncertainty estimated based on shot noise.}
    \label{fig:gravi_tstep}
\end{figure}

Previous studies \citep{Mace_2024, Palubski_2024} indicated that one may need to be careful in choosing the time step for simulations of SIDM halos to ensure accurate results. It has been claimed that depending on the set-up, simulations of the SIDM halo collapse require more stringent settings for the time step criteria than in common CDM simulations. This has been particularly expressed in terms of choosing smaller values for the time step parameter of the gravitational time step criterion.

In our simulations, we find that the value of the time step parameter $\eta$ has no significant influence on the results for the range that we have tested as shown in Fig.~\ref{fig:gravi_tstep}.
We note that $\Delta t_\mathrm{grav} \propto \sqrt{\eta}$ \citep[eq.~34 by][]{Springel_2005}.
In conclusion, we do not need a stricter time step criterion for gravity compared to the CDM simulations. However, we also note that we use at the same time our time step criterion for SIDM \citep{Fischer_2024a}. It becomes relevant at the late stage of the evolution ($t \gtrsim 4.4 \, \mathrm{Gyr}$) for the particles in the inner region. Here, it results in smaller time steps than solely based on the gravitational time step criterion. Moreover, we want to point out that it might be important how exactly the SIDM time step criterion is formulated \citep[see appendix~B by][]{Fischer_2024a}.

Furthermore, we want to point out that we have simulated a set-up for which the impact of the self-interactions is much stronger than numerical errors arising from solving for gravity. When simulating a DM halo over many more gravitational timescales, it could be that the desired accuracy can only be achieved when choosing a smaller time step, i.e.\ a smaller value for $\eta$ \citep[see also the discussion about the gravitational relaxation in Sect.~4 by][]{Fischer_2024b}.
However, if the numerical integration errors become significantly large, the collapse time of an SIDM halo simulation may no longer be inversely proportional to the cross-section.
This might be in line with the results \cite{Mace_2024} obtained for their simulations of small cross-sections that they have run for long time spans.

\section{Effective cross-section of the isolated halo}
\label{sec:sigma_eff_iso}

\begin{figure}
    \centering
    \includegraphics[width=\columnwidth]{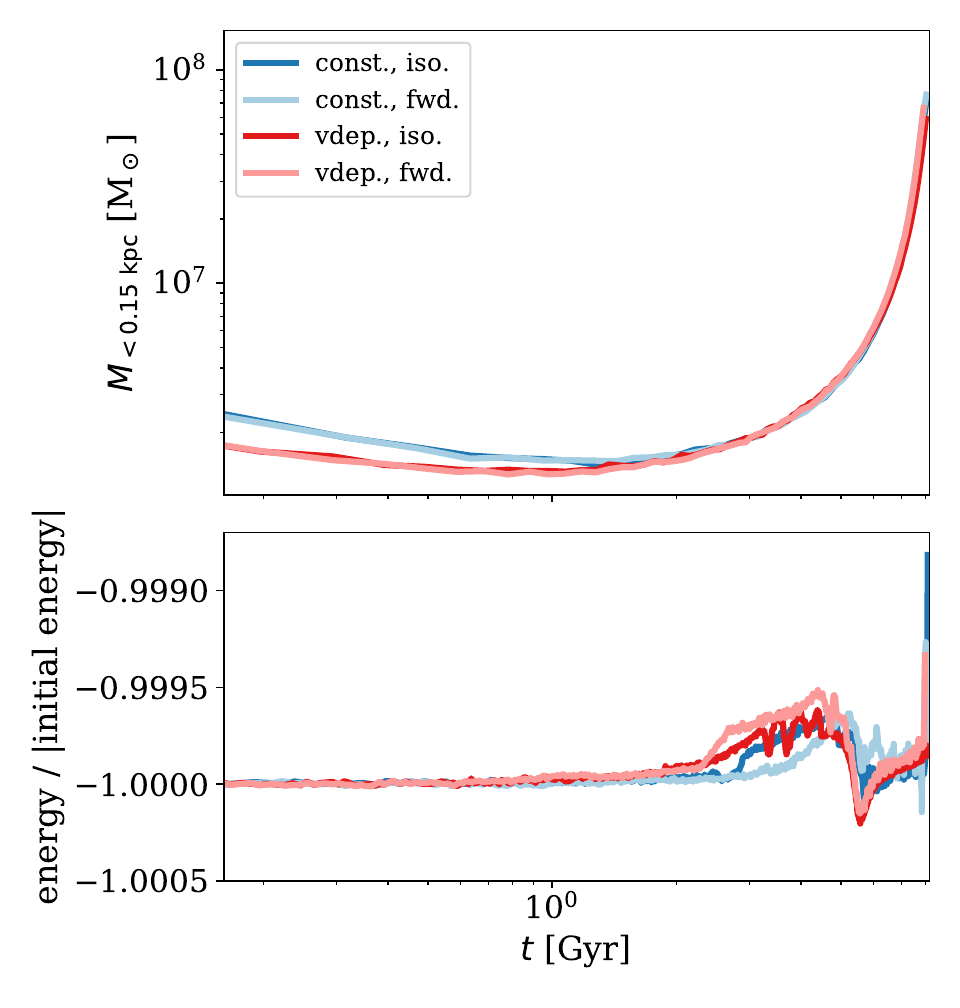}
    \caption{Enclosed mass and energy conservation as a function of time. We show the same simulations as in Fig.~\ref{fig:vdep+angle} (O, P, R, and S in Table~\ref{tab:sim_para}). The upper panel gives the mass within a radius of 0.15~kpc and the lower panels shows how well the total energy is conserved. The time for the simulations with a velocity-independent cross-section is rescaled to match the collapse phase of the simulations with velocity-dependent interactions.}
    \label{fig:rescaled}
\end{figure}

In this appendix, we discuss the differences between the velocity-independent and dependent runs for the isolated halo, as shown in Sect.~\ref{sec:angular_and_velocity_dep_iso}. In addition to Fig.~\ref{fig:vdep+angle}, we display the mass enclosed within 0.15~kpc as a function of time in Fig.~\ref{fig:rescaled}. We choose a larger radius to reduce the uncertainty in the measured mass. Importantly, we rescale the time axis of the velocity-independent runs to match the collapse phase of the velocity-dependent simulations. It becomes visible that there is a qualitative difference between the simulations, given that the core-expansion phase cannot be matched simultaneously. We note that the collapse timescale of the halo does not scale inversely proportional with the cross-section. This is because the halo is partially in the short-mean-free-path regime. If we nevertheless assume this scaling, the rescaled version of the velocity-independent cases would correspond to a cross-section of $\sigma_\mathrm{V}/m= 49.6 \, \mathrm{cm}^2 \, \mathrm{g}^{-1}$.

The effective cross-section \citep{Yang_2022D}, $\sigma_\mathrm{eff}$ (see Eq.~\eqref{eq:effective_cross_section}) is a common way to map the effect of SIDM models with different velocity and angular dependences on each other \citep[see also][]{Yang_2023}.
It can be expressed as
\begin{equation} \label{eq:effective_cross_section}
    \sigma_\mathrm{eff} = \frac{\langle v^5 \sigma_\mathrm{V}(v) \rangle}{\langle v^5 \rangle} \,.
\end{equation}
The average is computed by integrating over a Maxwell-Boltzmann distribution set by a characteristic velocity dispersion.
For the $\sigma_0/m$ value as in Eq.~\eqref{eq:vel_dep}, we employed a value of 90\% of the maximum velocity dispersion of the initial NFW halo for the characteristic velocity-dispersion, which is $17.28 \, \mathrm{km} \, \mathrm{s}^{-1}$, and found the corresponding effective cross-section is $80 \, \mathrm{cm}^2 \, \mathrm{g}^{-1}$, which is 61\% larger than the value from the rescaling in Fig.~\ref{fig:rescaled}, as discussed above.\footnote{In \cite{Yang_2022D}, the characteristic velocity dispersion is taken to be $0.64 \, v_\mathrm{max}$, where $v_\mathrm{max}$ is the maximum circular velocity of the initial NFW halo.}
For comparison, $w=20 \, \mathrm{km} \, \mathrm{s}^{-1}$ in the velocity-dependent models.
Moreover, $\sigma_\mathrm{V}$ denotes the viscosity cross-section normalised to match the total cross-section for isotropic scattering (Eq.~\eqref{eq:viscosity_cross_section}).

\begin{figure}
    \centering
    \includegraphics[width=\columnwidth]{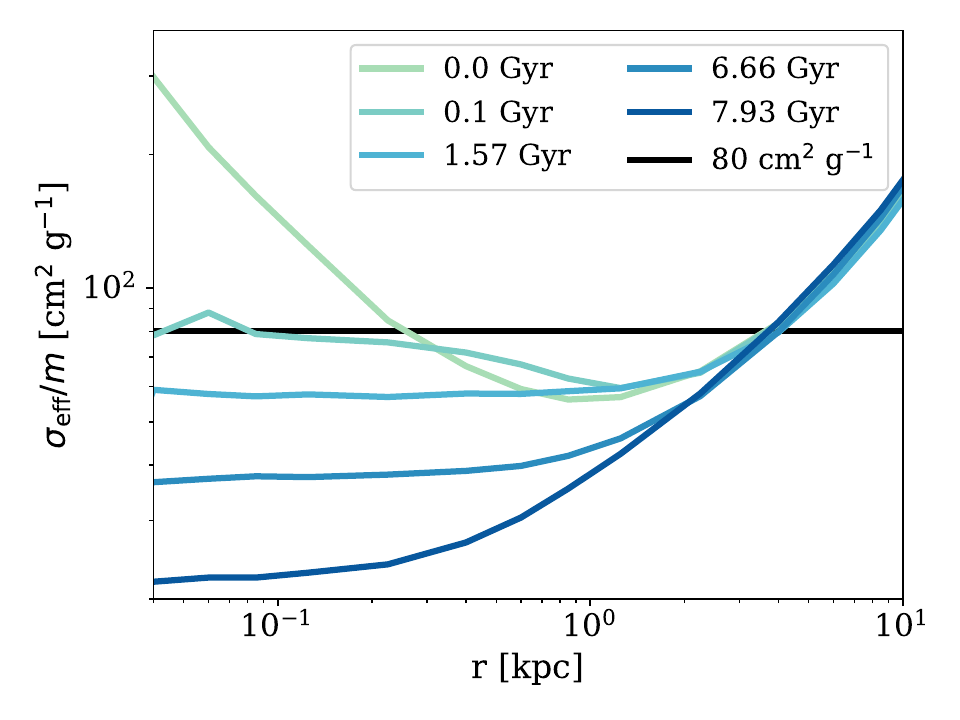}
    \caption{Effective cross-section as a function of radius. For the isolated halo evolved with an isotropic velocity-independent cross-section (simulation R according to Table~\ref{tab:sim_para}), we show the effective cross-section for several times as indicated in the legend. In addition, we display the cross-section for the velocity-independent cases (black line).}
    \label{fig:sigma_eff_iso}
\end{figure}

To allow for a better understanding of the difference between velocity-dependent and velocity-independent cases, we compute the effective cross-section (Eq.~\eqref{eq:effective_cross_section}) as a function of radius. Here, we do not use a single characteristic velocity dispersion, but instead use the value of the velocity dispersion at each radius from the simulation. In Fig.~\ref{fig:sigma_eff_iso}, we show the results for the simulation with isotropic scattering (run R according to Table~\ref{tab:sim_para}).
We can see that the effective cross-section initially is large at small and large radii but low for the intermediate range. At the early stage of core expansion, the constant cross-section $80 \, \mathrm{cm}^2 \, \mathrm{g}^{-1}$ is a reasonable approximation for the effective cross-section.
However, in later stages, the effective cross-section decreases a lot, which is a consequence of the increasing velocity dispersion and is in line with the long collapse time that we find in Fig.~\ref{fig:vdep+angle}. This continuous decrease during the collapse may limit the usability of the effective cross-section using a single characteristic velocity dispersion to map the effect that different SIDM models have on the evolution of an isolated halo on each other.
For the velocity-dependent models, we consider $w = 20 \, \mathrm{km} \, \mathrm{s}^{-1}$, which is comparable to the characteristic velocity dispersion of the initial halo $17.28 \, \mathrm{km} \, \mathrm{s}^{-1}$. Thus, the effective cross-section is sensitive to the change of the velocity dispersion in the halo during the core collapse. For models where $w$ is larger than the characteristic velocity dispersion, we expect that the mapping approach should work better. We will leave detailed studies on this topic for future work.

\section{Fit with King model}
\label{sec:king_fit}

\begin{figure}
    \centering
    \includegraphics[width=\columnwidth]{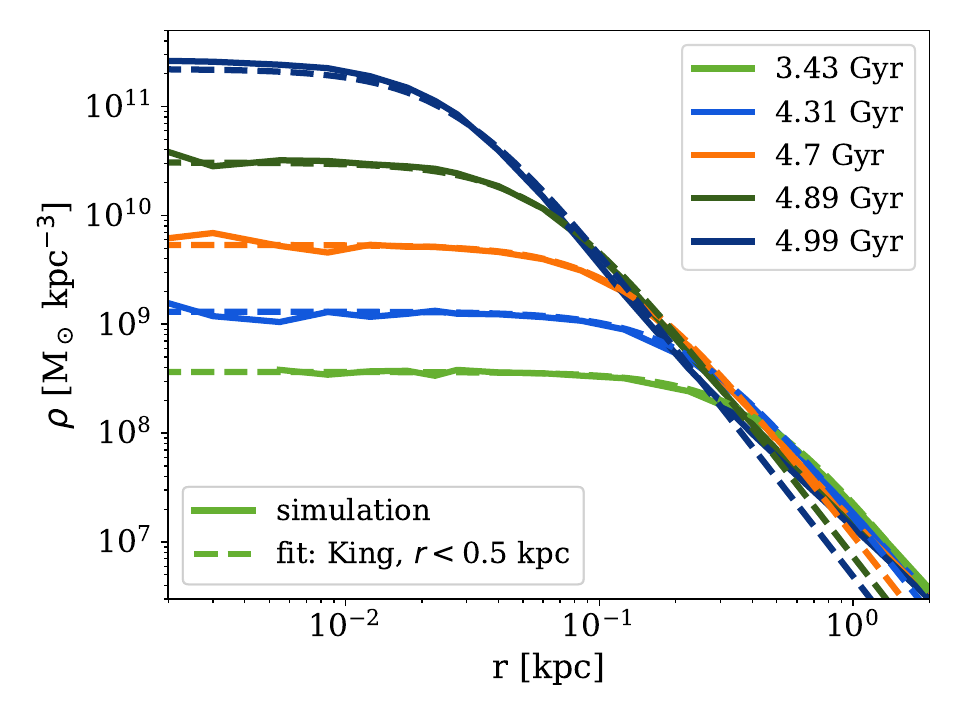}
    \caption{Density as a function of radius. Simulation results and a fit using the King model (Eq.~\eqref{eq:king}) are shown for different times of the late collapse phase. The results are for the isolated simulation with a resolution of $N=5 \times 10^7$ particles (simulation Q of Table~\ref{tab:sim_para}).}
    \label{fig:fit_king}
\end{figure}

\begin{figure}
    \centering
    \includegraphics[width=\columnwidth]{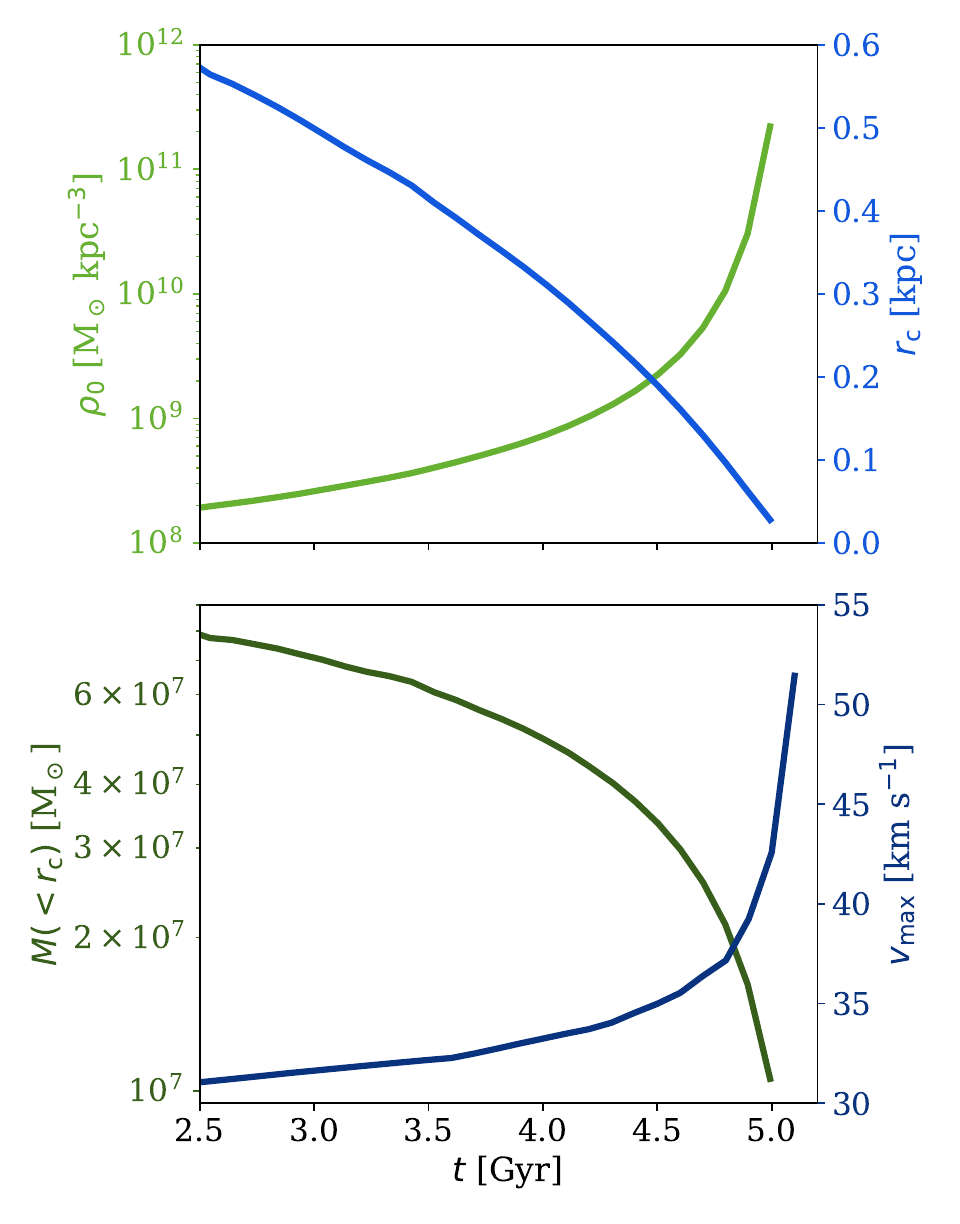}
    \caption{Parameters of fitted King model and further quantities as a function of time. The parameters that we determine by fitting a King model (Eq.~\eqref{eq:king}) to our highest-resolved simulation (Q in Table~\ref{tab:sim_para}) are shown in the top panel as a function of time. The bottom panel gives the mass enclosed in $r_\mathrm{c}$ according to the fitted model as well as the maximum circular velocity directly computed from the simulation data.}
    \label{fig:fit_king_param}
\end{figure}

In this appendix, we fit our highest-resolved simulation for the halo in isolation (simulation Q according to Table~\ref{tab:sim_para}) with the King model (Eq.~\eqref{eq:king}). As for the satellite halo in Sect.~\ref{sec:king_fit_sat}, we use a limited radial range only, $r<0.5 \, \mathrm{kpc}$.

In Fig.~\ref{fig:fit_king}, we give the density profile together with the fitted King model. The King model appears to provide a reasonable description of the central region of a collapsing SIDM halo.
However, it does not describe the outer regions well and may also eventually fail for the inner region at even later stages. 
Nevertheless, it may work well enough to fit observational data as discussed by \cite{Zhang_2025}.
We note that the central density of the fit, $\rho_0$, rises super-exponentially with time, and at the same time, the mass within $r_\mathrm{c}$ is decreasing as a function of time as we show in more detail below.

We display the parameters of the King model (Eq.~\eqref{eq:king}) that we obtained by fitting to our highest-resolved simulation in Fig.~\ref{fig:fit_king_param}.
It is visible that the central density quickly rises while the core size decreases.
Moreover, we show the enclosed mass within $r_\mathrm{c}$ and the maximum circular velocity ($v_\mathrm{vcirc} = \sqrt{\mathrm{G}\,M/r}$) as a function of time in the bottom panel.
Here, we can see that the core mass is quickly decreasing during the gravothermal collapse.

\FloatBarrier

\section{Orbit of the satellite and tidal evolution}
\label{sec:orbit_and_tidal_evolution}
\begin{figure}
    \centering
    \includegraphics[width=\columnwidth]{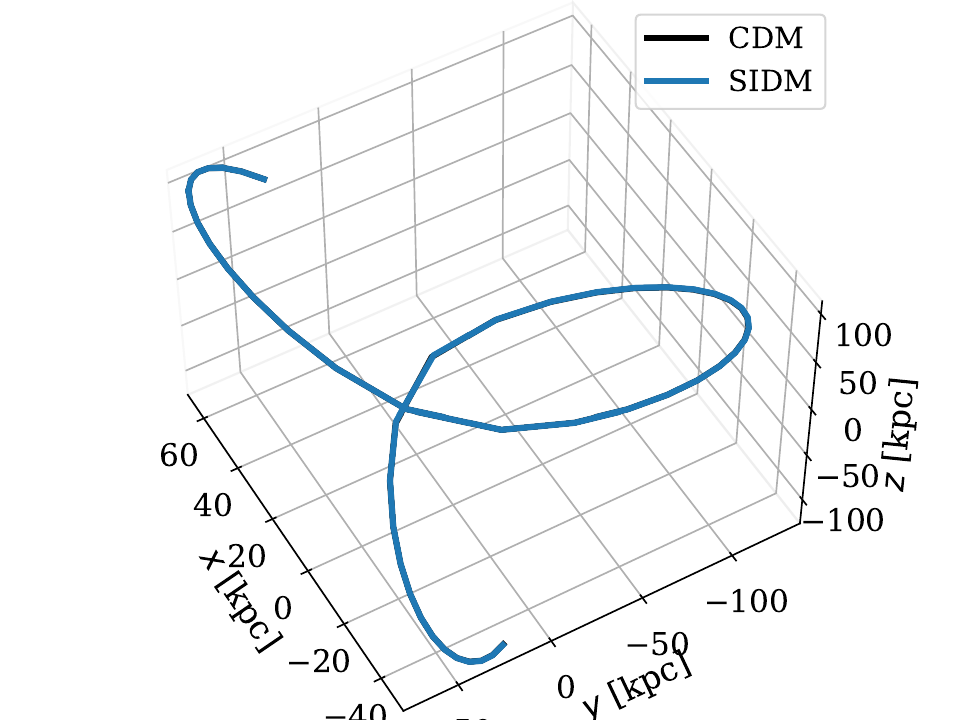}
    \caption{Trajectory of the satellite. The three-dimensional position of the subhalo for a CDM and an SIDM simulation is shown. The corresponding simulations are T and W according to Table~\ref{tab:sim_para}, where their parameters are given.}
    \label{fig:3dorbit}
\end{figure}

\begin{figure}
    \centering
    \includegraphics[width=\columnwidth]{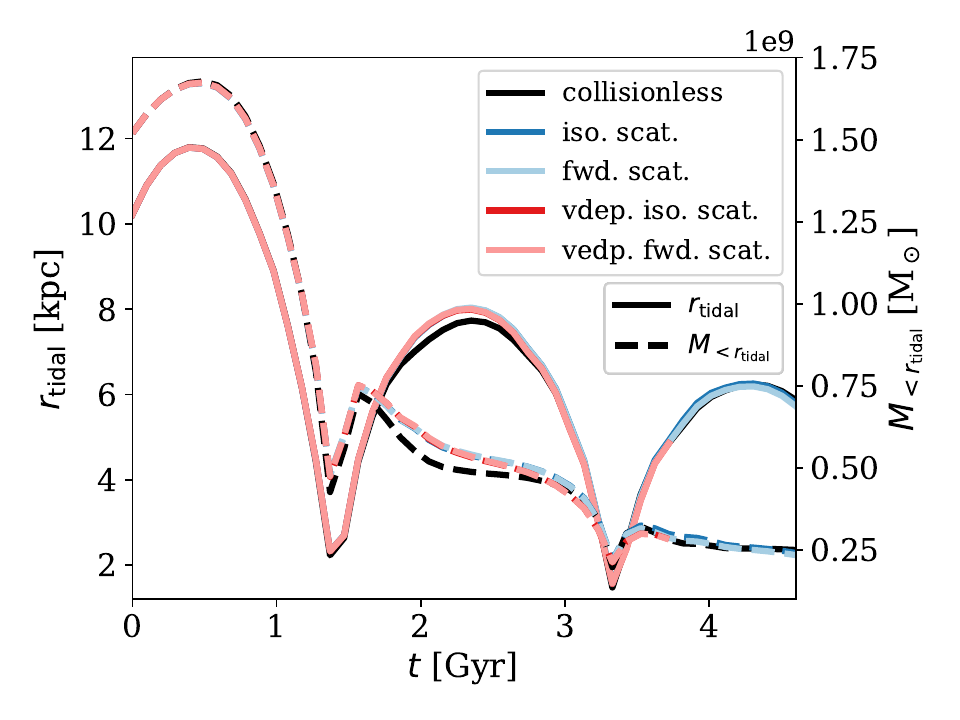}
    \caption{Tidal radius and its enclosed mass as a function of time. For the satellite halo, evolved with different DM models, the tidal radius (solid lines) according to Eq.~\eqref{eq:tidal_radius} and the mass within it (dashed lines) are shown. All simulations shown here share a resolution of $N=10^7$ particles and a gravitational softening length of $\epsilon=1.0 \, \mathrm{kpc}$. Further information on the simulations T, W, X, Y, and Z is provided in Table~\ref{tab:sim_para}.}
    \label{fig:tidal_evolution}
\end{figure}

To give an overview, we show the trajectory of the satellite in Fig.~\ref{fig:3dorbit}. The results for a CDM and an SIDM simulation are shown. The two orbits are almost identical. This is also related to the fact that we neglect dynamical friction and the non-gravitational interaction between the satellite's DM particle with the host's DM.

Next, we investigate the tidal evolution of the satellite for the DM models that we have simulated.
Here, we compute the tidal radius, $r_\mathrm{tidal}$, of the satellite halo. For this, we implicitly define it as
\begin{equation} \label{eq:tidal_radius}
    \frac{\mathrm{G} \, M_\mathrm{sat}(<r_\mathrm{tidal})}{r_\mathrm{tidal}^2} = \left|\nabla \Phi_\mathrm{host}(\mathbf{x} + r_\mathrm{tidal} \, \mathbf{n}) - \nabla \Phi_\mathrm{host}(\mathbf{x} - r_\mathrm{tidal} \, \mathbf{n})\right| \,.
\end{equation}
Here, $M_\mathrm{sat}(<r_\mathrm{tidal})$ denotes the mass of the satellite halo that is enclosed within $r_\mathrm{tidal}$ and $\Phi_\mathrm{host}$ is the gravitational potential of the host, which in our case is given analytically as described in Sect.~\ref{sec:satellite_setup}.

In Fig.~\ref{fig:tidal_evolution}, we display the tidal radius as a function of time for various DM models and show its enclosed mass. It is visible that all DM models evolve fairly similarly, especially the SIDM models. This is related to the fact that the tidal radius is typically relatively large compared to the initial scale radius of the satellite halo ($r_\mathrm{s} = 1.28 \, \mathrm{kpc}$).

\FloatBarrier

\section{Effective cross-section of the satellite}
\label{sec:sigma_eff_sat}
\begin{figure}
    \centering
    \includegraphics[width=\columnwidth]{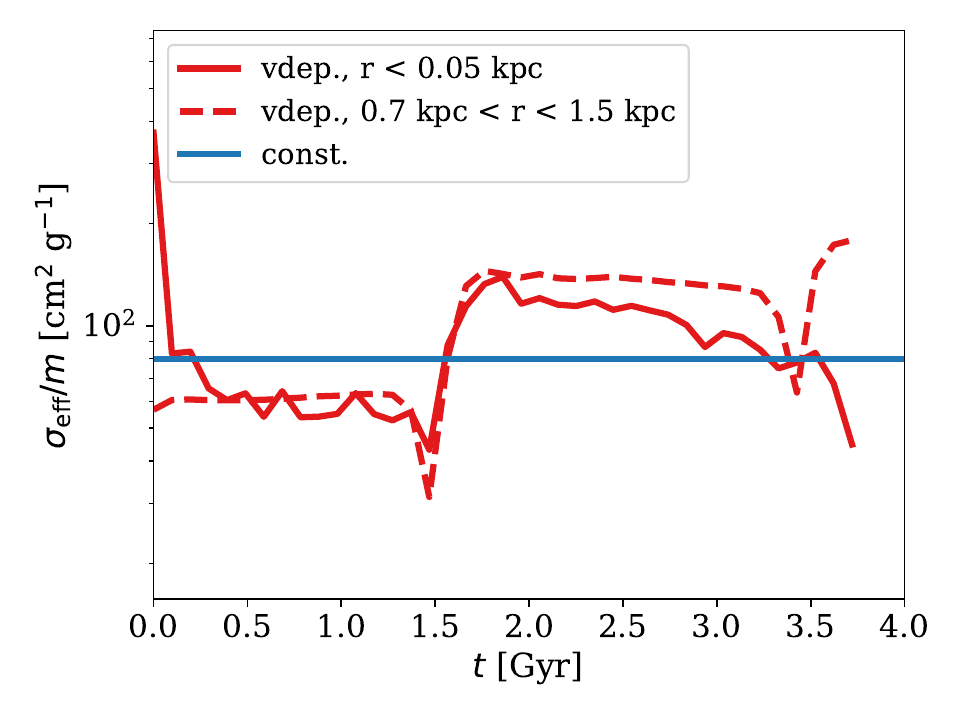}
    \caption{Effective cross-section as a function of time. We compute the effective cross-section for different radial ranges from the average velocity dispersion using the satellite simulation for an isotropic velocity-dependent cross-section (simulation Y of Table~\ref{tab:sim_para}). Moreover, we show the velocity-independent case, which does not have any radial dependence.}
    \label{fig:sigma_eff_sat}
\end{figure}
In this appendix, we study the effective cross-section given by Eq.~\eqref{eq:effective_cross_section} for our simulation of a satellite halo begin subject to an isotropic velocity-dependent cross-section. We compute the results given in Fig.~\ref{fig:sigma_eff_sat} using simulation Y as specified in Table~\ref{tab:sim_para}. The effective cross-section requires a characteristic velocity dispersion. Here, we use the average velocity dispersion within two different radial ranges, namely $r<0.05\,\mathrm{kpc}$ and $0.7\,\mathrm{kpc} < r < 1.5\,\mathrm{kpc}$. In addition, we show the case of the constant cross-section of $80\,\mathrm{cm}^2 \, \mathrm{g}^{-1}$ as we use it for the simulations for a velocity-independent cross-section.

It is visible in Fig.~\ref{fig:sigma_eff_sat}, that the effective cross-section for the velocity-dependent case is larger in the centre of the halo (red, solid) than for the constant case (blue). This leads to an enhanced core formation. However, later on, the effective cross-section drops below $80 \, \mathrm{cm}^2 \, \mathrm{g}^{-1}$. At about the first pericentre passage ($\approx 1.5 \, \mathrm{Gyr}$) tidal stripping is effective and reduces the velocity dispersion of the satellite halo, as discussed in Sect.~\ref{sec:angular_and_velocity_dep_sat}. As a consequence, the effective cross-section sharply rises for the velocity-dependent case. This is the case at small (dotted) and large (dashed) radii. Subsequently, the effective cross-section at small radii starts to decline because of the increase in velocity dispersion due to the gravothermal collapse. At larger radii, the impact of the second pericentre passage is visible, increasing the effective cross-section for later stages.

\section{Highest-resolved simulation}
\label{sec:high_res}

\begin{figure}
    \centering
    \includegraphics[width=\columnwidth]{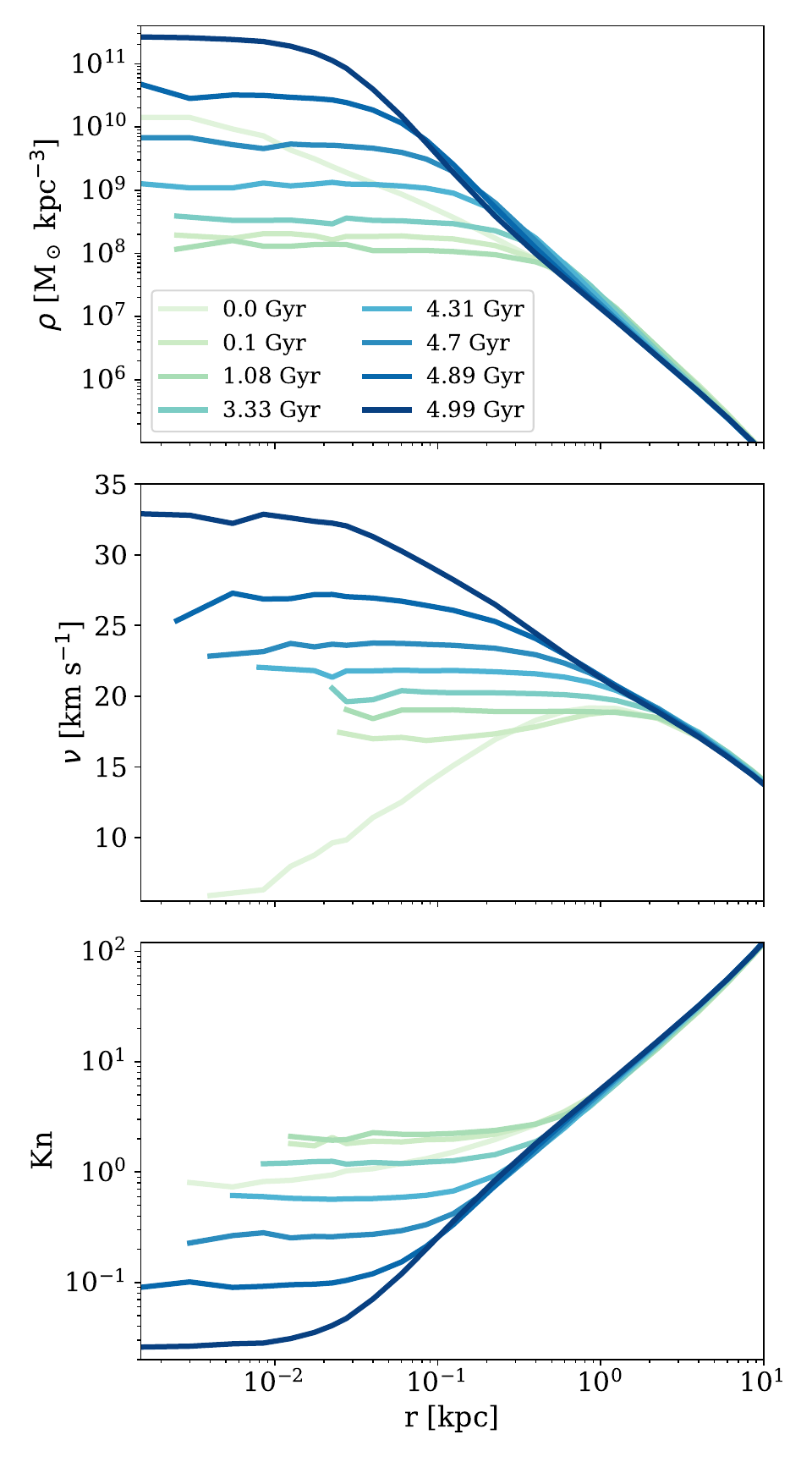}
    \caption{Density, velocity dispersion, and Knudsen number as a function of radius. We show the results for our most highly resolved halo (simulation Q of Table~\ref{tab:sim_para}) at various stages of the gravothermal evolution.}
    \label{fig:profiles}
\end{figure}

In this Appendix we provide additional information for our highest-resolved simulation evolved in isolation (simulation Q in Table~\ref{tab:sim_para}).
Figure~\ref{fig:profiles} gives the density (top panel), the one-dimensional velocity dispersion (middle panel), and the Knudsen number (bottom panel) as a function of radius for several times. Here, the gravothermal evolution driven by a velocity-independent cross-section including the core expansion and collapse phase is well visible.
\end{appendix}

\end{document}